\PassOptionsToPackage{splitbox=false}{adjustbox}
\documentclass[twocolumn,twocolappendix,tighten,times,astrosymb]{aastex631}

\usepackage{amsmath}
\usepackage{booktabs}
\usepackage{soul}
\usepackage{microtype}
\usepackage[english]{babel}
\usepackage{cancel}
\usepackage{soul}

\makeatletter
\@ifundefined{splitbox}{}{%

}
\makeatother

\usepackage{adjustbox}

\hypersetup{linkcolor=blue,citecolor=blue,filecolor=blue,urlcolor=blue}

\newcommand{\be}{\begin{equation}}
\newcommand{\ee}{\end{equation}}

\newcommand{\plenum}{$\texttt{PLE} \nu \texttt{M}$}

\long\def\exclude#1{}

\shorttitle{Neutrinos from AGN coronae}
\shortauthors{}
\begin{document}

\title{
\large Neutrinos and gamma rays from Seyfert galaxies constrain the properties of coronal turbulence
}

\correspondingauthor{federico.testagrossa@desy.de}
\author[0009-0000-9401-1971]{Federico Testagrossa}
\affiliation{Deutsches Elektronen-Synchrotron DESY, Platanenallee 6, 15738 Zeuthen, Germany}
\author[0000-0003-4927-9850]{Damiano F. G. Fiorillo}
\affiliation{Istituto Nazionale di Fisica Nucleare (INFN), Sezione di Napoli, Complesso Universitario di Monte Sant'Angelo, Via Cintia, 80126 Napoli, Italy}
\affiliation{Deutsches Elektronen-Synchrotron DESY, Platanenallee 6, 15738 Zeuthen, Germany}
\author[0000-0001-8822-8031]{Luca Comisso}
\affiliation{Department of Physics, Columbia University, New York, NY 10027, USA}
\affiliation{Department of Astronomy, Columbia University, New York, NY 10027, USA}
\author[0000-0003-0543-0467]{Enrico Peretti}
\affiliation{INAF - Astrophysical Observatory of Arcetri, Largo E. Fermi 5,
50125 Florence, Italy}
\affiliation{Université Paris Cité, CNRS, Astroparticule et Cosmologie, 10 Rue Alice Domon et Léonie Duquet, F-75013 Paris, France}
\author[0000-0001-6640-0179]{Maria Petropoulou}
\affiliation{Department of Physics, National and Kapodistrian University of Athens, University Campus Zografos, GR 15784, Athens, Greece }
\affiliation{Institute of Accelerating Systems \& Applications, University Campus Zografos, Athens, Greece}
\author[0000-0002-1227-2754]{Lorenzo Sironi}
\affiliation{Department of Astronomy and Columbia Astrophysics 
Laboratory, Columbia University, New York, NY 10027, USA}
\affiliation{Center for Computational Astrophysics, Flatiron Institute, 162 5th Avenue, New York, NY 10010, USA}

\begin{abstract}
{The TeV neutrino signal observed by IceCube from the active galactic nucleus (AGN) NGC 1068 can probe its innermost coronal regions. If these neutrinos originate from hadrons accelerated within a magnetized turbulent corona, their intensity and spectrum depend on the turbulent magnetic field strength and turbulence coherence scale. The gamma rays accompanying neutrino production are absorbed in this optically thick environment, in a way that depends sensitively on the size of the corona. By a joint fit of the IceCube and Fermi-LAT observations, we translate the multimessenger signal from NGC~1068 and the tentative signal from NGC~7469 into quantitative constraints on coronal properties. NGC~1068, with a significant TeV neutrino excess, favors a compact, strongly magnetized corona with a large turbulence coherence length relative to the coronal size. NGC~7469, with two $\sim 100$~TeV neutrino events, points instead to a somewhat larger corona with much smaller coherence length and high magnetization, but a very small fraction of energy in non-thermal protons. We obtain the diffuse flux from a population of Seyfert galaxies identical to either NGC~1068 or NGC~7469. Finally, we consider a third scenario, motivated by the spectral break observed in the diffuse neutrino flux at tens of TeV, with coronal properties intermediate between the two point-source-inspired models. To enable detailed comparisons with the IceCube and electromagnetic observations, we release our model predictions in a GitHub repository.}

\end{abstract}


\section{Introduction}

High-energy neutrinos are unique messengers among astrophysical particles. Thanks to their extremely weak interactions, they can escape from dense regions where charged particles are deflected and cooled, and gamma rays are absorbed. Their emission is primarily driven by hadronic processes, so they carry direct information about non-thermal hadrons in cosmic accelerators.
On the other hand, their weak coupling also makes them notoriously hard to detect, hindering unambiguous associations with their astrophysical sources. For almost a decade after their first discovery~\citep{IceCube2013}, high-energy neutrinos came mostly in the form of an unresolved diffuse flux spanning roughly 10~TeV to 10~PeV.

More recently, improved statistics have enabled the identification of directional excesses linked to specific candidates. The brightest hotspot in the neutrino sky corresponds to the Seyfert galaxy NGC 1068~\citep{IceCube:2019cia,IceCube:2022der}. Its active galactic nucleus (AGN) -- the compact, accreting region around the central supermassive black hole (SMBH) -- is the most plausible origin of the neutrino emission~\citep{Padovani:2024ibi}. A striking clue is the absence of comparably strong TeV gamma-ray emission~\citep{MAGIC-UL-NGC1068}, suggesting that neutrinos emerge from a radiatively compact region where gamma rays are attenuated and reprocessed to lower energies \citep{Murase_2020, Murase_2022ApJL}. The overall significance of the NGC 1068 signal is around $4\sigma$, varying slightly with the adopted source catalog~\citep{IceCube:2022der,IceCube24_Seyfert,Abbasi:2025tas}.

A natural site for such neutrino emission is the AGN corona -- a compact, magnetized region whose extent is typically inferred to be a few tens of gravitational radii from the SMBH~\citep[see e.g.][and references therein]{Cackett2021}. The corona is invoked to explain the ubiquitous hard X-ray spectra of AGN via Comptonization by energetic electrons. Within this environment, the photon field is dense enough to suppress gamma rays above tens of MeV, while allowing neutrinos produced in hadronic interactions to escape. Hence, neutrino observations offer a direct probe of the magnetic field and non-thermal particle content of AGN coronae, otherwise inaccessible to electromagnetic diagnostics. Indeed, IceCube has reported growing evidence for neutrino associations with other Seyfert galaxies similar to NGC 1068, albeit at lower significance~\citep{Neronov_Seyfert,IceCube24_Seyfert,Abbasi:2025tas}, opening the prospect of testing the diversity of AGN inner regions through neutrinos.

The key open question is how non-thermal protons are accelerated in the corona to the energies required by the IceCube detections. Early works focused on diffusive shock acceleration~\citep{Inoue_2020} and gyroresonant stochastic acceleration in weak magnetized turbulence~\citep{Dermer:1995ju,Murase:2019vdl}. More recent scenarios include direct acceleration within reconnection regions~\citep{Fiorillo:2023dts,Karavola:2024uui, Karavola:2026rpg, Passos-Reis:2026mmy} and proton pre-acceleration in magnetic reconnection layers followed by stochastic re-acceleration~\citep{Mbarek:2023yeq}. The stochastic-acceleration picture has since evolved: in~\cite{Fiorillo:2024akm}, we showed -- motivated by recent particle-in-cell (PIC) simulations~\citep{Comisso:2019frj,Bresci:2022awc,ComissoFM2024} -- that turbulent energization in the corona must be treated as non-resonant. Using PIC-based constraints, we inferred that, in order to power the NGC~1068 neutrino flux, the corona must be strongly magnetized, with a turbulent magnetic energy density comparable to the rest-mass energy density and well above the thermal one. Extending this framework to other Seyfert galaxies, ~\cite{Fiorillo:2025ehn} found that a consistent set of coronal parameters can reproduce the diffuse neutrino background between 10–100 TeV.
The spectral break in the diffuse flux around 20~TeV, recently observed by IceCube~\citep{IceCube:2025tgp}, was directly explained as the maximal neutrino energy produced in the corona. Non-resonant stochastic acceleration has also been considered in~\cite{Lemoine:2025roa,Yang:2025lmb}, which emphasized a possible feedback of non-thermal protons on the turbulent cascade. 

In this work, we advance our scenario of a strongly magnetized turbulent corona, fitting the combined neutrino and electromagnetic emission to the data of candidate neutrino-emitting coronae to pinpoint their source properties. Neutrino observations are crucial for this task, as gamma rays alone can at most provide upper bounds on the coronal size, but are insensitive to the magnetic properties of the corona. These properties are encoded in four parameters: the turbulent magnetization $\sigma_{\rm tur}=2U_{B, \, \rm tur}/U_{\rm rest}$, namely the ratio between the  energy density of the turbulent magnetic field $U_{B, \, \rm tur}$ and the rest-mass energy density of the plasma $U_{\rm rest}$; the coronal radius $R$; the coherence length of the turbulence $\ell$, measured by the dimensionless parameter $\eta=\ell/R$; and the fraction of magnetically dissipated energy attained by non-thermal protons $\mathcal{F}_p$. Electromagnetic observations alone are unable to probe the magnetic properties of the corona $\sigma_{\rm tur}$, $\eta$, and $\mathcal{F}_p$. Thus we perform a combined analysis of the IceCube neutrino observations and Fermi-LAT measurements of the GeV gamma-ray flux from 
all of the reported Seyfert galaxies, determining the allowed parameter space for each of them. We also discuss the implications of the results for the diffuse neutrino flux. Our work comes with an associated GitHub release~\citep{GitRepo}, where we collect our multimessenger predictions for the parameterized turbulent model for galaxies with varying X-ray luminosity, enabling direct comparisons with observations from Seyfert galaxies beyond simple phenomenological approaches. 

After reviewing in Sec.~\ref{sec:coronal_model} the model of a strongly magnetized turbulent corona firstly introduced in \cite{Fiorillo:2024akm}, in Sec.~\ref{sec:analysis} we describe our statistical analysis of the gamma-ray and neutrino emission from Seyfert galaxies, whose results are presented in Sec.~\ref{sec:fit_coronal}. In Sec.~\ref{sec:back_of_the_envelope} we present back-of-the-envelope estimates of the coronal properties, completely consistent with the more detailed analysis. In Sec.~\ref{sec:diffuse}, we discuss the implications for the diffuse neutrino flux. Finally, in Sec.~\ref{sec:discussion}, we discuss the more general implications of our results.

\section{Coronal model}\label{sec:coronal_model}

Neutrino and photon emission happens within the corona, whose properties are largely unknown. Our only assumption is therefore that neutrinos and gamma rays are powered by the turbulent dissipation of energy within a compact region close to the central black hole. Following~\cite{Fiorillo:2024akm,Fiorillo:2025ehn}, we model this region as a sphere with radius $R$ \footnote{A non-spherical shape might somewhat alter the results by factors of order unity} typically of the order of $R\simeq 10-100\,r_g$, where $r_g=GM/c^2$ is the gravitational radius of the central SMBH with mass $M$; here $G$ is the gravitational constant. We express our results in terms of the scaling parameter $M=M_7\,10^7\,M_\odot$ for the SMBH mass. The precise value of the coronal radius $R$ will be used as a fit parameter to ensure that the corona is compact enough to be opaque to gamma rays.

\subsection{Coronal properties -- electrons, thermal protons and magnetic fields}

Within the corona, a population of energetic electrons is responsible for the production of X-rays via Comptonization of photons emitted from the accretion disk; we do not model the production of X-rays, but rather consider them as a fixed photon field. The electron density is inferred from the condition that the optical thickness for Comptonization $\tau_T=n_e R \sigma_T\sim 0.5$, where $\sigma_T$ is the Thomson cross section. The value of $\tau_T \sim 0.5$ is typically inferred for AGN coronae~\citep[e.g.][]{2018MNRAS.480.1819R}. Hence, the electron number density is
\begin{equation}
    n_e\simeq 2.5\times 10^{10}\;\mathrm{cm}^{-3}\; \frac{20 \, r_g}{R M_7}\,.
    \label{eq:pair_density}
\end{equation}
In reconnection models \citep{Fiorillo:2024akm, Karavola:2024uui}, where the corona is associated with the current sheets that form in the magnetosphere of the SMBH, a few $r_g$ from the event horizon, the corona is expected to be dominated by electron-positron pairs $n_e \gg n_p$.

In this work we focus on a turbulent corona model, which is associated with a larger corona ($R\ge 10\,r_g$) surrounding the SMBH and sustained by the accretion flow. Assuming a pure hydrogen plasma and quasi-neutrality, we have $n_p\simeq n_e$ in the corona; therefore, the rest-mass energy density of the plasma is
\begin{equation}
    U_{\rm rest}\simeq n_pm_pc^2\simeq 3.9\times 10^7\;\mathrm{erg/cm}^3\;\frac{20 r_g}{R M_7} \, ,
\end{equation}
where $m_e$ and $m_p$ are the electron and proton masses, respectively. 

We assume a stationary turbulent environment, with a global magnetic field $B$ and magnetic energy density $U_B=B^2/8\pi$, and an outer-scale turbulent magnetic field $\delta B$ whose intensity is measured by the dimensionless parameter $\eta_B=\delta B^2/B^2$, so that the turbulent magnetic energy density is $U_{B,\rm tur}=\eta_B B^2/8\pi$. It will be convenient to express the magnetic field strength in terms of another dimensionless parameter, namely the magnetization 
\begin{equation}\label{def:sigma}
    \sigma=\frac{2U_B}{U_{\mathrm{rest}}}\simeq\frac{B^2}{4\pi n_pm_pc^2},
\end{equation}
and the corresponding turbulent magnetization $\sigma_{\rm tur}=\sigma \eta_B$. We will consider values of $0.01\lesssim \sigma_{\rm tur}\lesssim1$, with magnetic energy density not exceeding the rest mass energy density of the corona; as we will see, the actual observations favor in fact $\sigma_{\rm tur}\sim 1$, which we denote as a strong-turbulence regime. The turbulent magnetization also determines the Alfvén velocity associated with the turbulent field $(v_A/c)^2=\sigma_{\rm tur}/(1+\sigma_{\rm tur})\simeq \min(\sigma_{\rm tur},1)$, so that large turbulent magnetizations are associated with relativistic magnetized plasmas with $v_A\sim c$. Finally, the strength of the magnetic field can also be measured in terms of the so-called plasma beta parameter $\beta$, relating the thermal pressure of the plasma $P_{\rm ther}$ to the magnetic pressure $P_B = U_B$. As in~\cite{Murase_2020}, we assume the proton temperature to be roughly of the order of the virial temperature $k_BT_p=GM m_p/3R$, so that the plasma beta parameter can be written as  
\begin{equation}\label{eq:beta_parameter}
    \beta=\frac{P_{\rm ther}}{P_B} = \frac{8\pi n_p k_B T_p}{B^2}=\frac{2r_g}{3R \sigma}.
\end{equation}
On the other hand, as we will see, since the magnetic energy is much larger than the gravitational energy density, we cannot exclude that protons get significantly hotter, so this only serves as a qualitative measure of the importance of magnetic energy.
The beta parameter associated with the turbulent fields is then $\beta_{\rm tur}=\beta/\eta_B$. Hereafter we will assume strong turbulence, i.e. the turbulent fluctuations of the magnetic field are of the same order of the global magnetic field, $\delta B\sim B$. Smaller fluctuation amplitudes would tighten the energetic requirements and reduce the energy budget available for neutrino production, which is already demanding, as discussed in \cite{Fiorillo:2024akm}. 
For this case we set $\eta_B = 1$ and $\sigma = \sigma_{\mathrm{tur}}$, although we will maintain the distinction between the two in our analytical expressions. The final parameter characterizing the magnetic field is its coherence length $\ell$, which is naturally measured in terms of the coronal radius $\ell=\eta R$.

A final quantity of great interest for the dynamics of the corona is the magnetic dissipation power $L_B$, i.e. the amount of energy that the magnetic field can dissipate. Assuming the corona to be magnetically powered, this rate must be large enough to accommodate X-ray emission and non-thermal proton acceleration. Assuming a reconnection rate $\eta_{\rm rec}\simeq 0.1$ in the reconnection layers embedded in the turbulence \citep{CS18,Comisso:2019frj} where most of the turbulent energy is ultimately dissipated, we use the expression deduced in~\cite{Fiorillo:2024akm}
\begin{equation}
    L_B=\frac{2\pi}{3}\frac{\eta_{\rm rec}}{\eta}\frac{c^3 \sigma_{\rm tur}^{3/2}}{\max(1,\sigma^{1/2})}n_p m_p R^2,
    \label{eq:B_luminosity}
\end{equation}
which roughly corresponds to assuming that a fraction $\eta_{\rm rec}$ of the turbulent magnetic energy density can be dissipated over an Alfvén timescale $t_A\sim \ell/v_A$.

\subsection{Non-thermal protons}

A small fraction of the thermal protons can be energized sufficiently as to enter the stochastic acceleration process in the magnetized turbulence. The typical timescale for stochastic proton acceleration inferred from PIC simulations is \citep{Comisso:2019frj}
\begin{equation}\label{eq:acceleration_rate}
    t_{\rm acc}\simeq \frac{10 \, \eta R}{\sigma_{\rm tur}c}, 
\end{equation}
which is energy independent, scales linearly with $\ell = \eta R$, and is inversely proportional to $\sigma_{\rm tur}$. As discussed in more detail in~\cite{Fiorillo:2024akm}, the acceleration is largely non-resonant and dominated by the interaction with large intermittent structures (see also~\cite{Lemoine:2022rpj,SC25}).
The energy distribution of the non-thermal protons is described by their phase-space distribution $f_p(p)$; since protons are ultra-relativistic\footnote{We do not deal here with the injection problem, which is discussed in more detail in~\cite{Fiorillo:2024akm}.}, their energy can be approximated as $E_p=pc$, and the energy distribution is therefore $dn_p/dE_p=4\pi p^2 f_p(E_p/c)/c$. The specific form of the energy distribution depends on the relative role of acceleration, escape, and cooling timescales, which we discuss below. We quantify the fraction of magnetically dissipated energy given to non-thermal protons per unit time with the dimensionless ratio $\mathcal{F}_p=L_p/L_B$, where $L_p$ is the energy given per unit time to the non-thermal proton population from the acceleration mechanism. As shown in~\cite{Fiorillo:2024akm}, this is defined as
\begin{equation}
    L_p=-4\pi c\frac{4\pi R^3}{3}\int \frac{p^4}{t_{\rm acc}}\frac{\partial f_p}{\partial p}dp;
\end{equation}
integrating by parts and substituting $t_{\rm acc}$ from Eq.~\ref{eq:acceleration_rate} we see that this can be expressed in terms of the total energy density of the non-thermal protons $U_{p}$ as
\begin{equation}
    L_p=\frac{8\pi c R^2 U_{p}\sigma_{\rm tur}}{15\eta}.
\end{equation}

The concrete proton spectrum, which in turn determines the coronal electromagnetic and neutrino radiation, depends also on their cooling losses and their escape rate from the acceleration region. We now discuss both these processes. 

A guaranteed, but subdominant, source of cooling is synchrotron radiation; the timescale for energy loss is
\begin{equation}
    t_{\rm synch}=\frac{3m_p^3c}{2\sigma_T m_e^2 E_p \sigma n_e}.
\end{equation}
A more important source of cooling is due to proton-proton ($pp$) collisions with the thermal protons with density $n_p$. We assume $n_p = n_e$, with the latter inferred from the Compton opacity argument of Eq.~\ref{eq:pair_density}. The $pp$ cooling timescale
\begin{equation}
    t_{pp}\simeq (n_p \sigma_{pp}(E_p)\kappa_p c)^{-1}
\end{equation}
depends on the scattering cross section $\sigma_{pp}(E_p)$ and the inelasticity $\kappa_p\simeq 0.5$.

The dominant cooling processes are due to proton-photon interaction in the dense target photon field of the AGN. These entail both Bethe-Heitler processes $p+\gamma\to p+e^++e^-$, with timescale $t_{\rm BH}$, and photohadronic scattering $p\gamma$, with timescale $t_{p\gamma}$. The cooling rate for protons associated with these two processes depends on the assumed target photon field; for each galaxy considered in this study, our chosen ambient photon field is discussed in Sec.~\ref{sec:analysis}. The timescales for energy losses from each of these processes are determined by the numerical code \texttt{AM}$^3$~\citep{Klinger:2023zzv}, which self-consistently treats also the radiation from hadrons.

Finally, the escape rate of protons from the coronal environment also plays a significant role in shaping their energy distribution. A guaranteed escape channel is proton diffusion, which we parameterize as~\citep{Fiorillo:2024akm,Fiorillo:2025ehn}
\begin{equation}
    t_{\rm esc}\simeq \frac{R}{c}\mathrm{max}\left[1,\frac{R}{\ell}\left(\frac{eB\ell}{E_p}\right)^{1/3}\right].
\end{equation}
Additional escape mechanisms associated with large-scale hydrodynamical motions have been discussed in the literature~\citep{Lemoine:2025roa}. We have considered the impact of hydrodynamical escape on the proton distribution in~\cite{Fiorillo:2025ehn}; however, as discussed there, we regard this as a more model-dependent possibility, since it requires specific assumptions on the coronal structure (e.g. a strong regular magnetic field may have the opposite effect of confining the fluid). For this reason, we do not include hydrodynamical escape as an additional escape channel in this work.

The overall proton dynamics is set by a competition between acceleration processes, with typical timescale $t_{\rm acc}$; cooling processes, with a combined rate
\begin{equation}
    t_{\rm cool}^{-1}=t_{pp}^{-1}+t_{p\gamma}^{-1}+t_{\rm BH}^{-1}+t_{\rm synch}^{-1},
\end{equation}
and escape, with timescale $t_{\rm esc}$. The evolution of protons is ruled by a transport equation for their phase-space distribution $f_p$, 
\begin{equation}\label{eq:transport}
    \partial_t f_p=p^{-2}\partial_p\left(\frac{p^4\partial_p f_p}{t_{\rm acc}}+\frac{p^3 f_p}{t_{\rm cool}}\right)-\frac{f_p}{t_{\rm esc}}+q_p,
\end{equation}
where $q_p$ is an injection term at low energies, whose specific form is irrelevant at high energies where it vanishes (this can be seen e.g. in Fig.~1 of \cite{Yuan:2025ctq}).

In our model the corona is described by the four parameters $R$, $\sigma_{\mathrm{tur}}$, $\eta$, $\mathcal{F}_p$, and by the target photon field produced by the AGN of the host galaxy. These elements determine the timescales in Eq.~\ref{eq:transport}. Since the radiation produced by non-thermal protons will be computed with the numerical code $\texttt{AM}^3$ \citep{Klinger:2023zzv}, for consistency we use   $\texttt{AM}^3$ also to compute the cooling timescales. Once those are specified, the equation is well-defined and admits a steady solution with $\partial_t f_p=0$. Thus, for each galaxy analyzed in Sec.~\ref{sec:analysis}, we  solve the stationary transport equation following the same numerical strategy as \cite{Fiorillo:2024akm}. 

\subsection{Energetics of the coronal environment}\label{sec:energetics}

\begin{table*}[t]
\centering
\begin{tabular}{|l|c|c|c|c|c|}
\hline
\textbf{Galaxy} & SED & $L_X^{2-10\,\mathrm{keV}}$ $[\text{erg s}^{-1}]$ & $d_L$ [Mpc] & $M_{\rm BH}$ [$M_\odot$] & Fermi-LAT \\
\hline
NGC 1068 & (a) & $3\times10^{43}$ (b) & 10.1 (b) & $0.67\times10^7$\,(b) & (c) \\
\hline
NGC 7469 & Fig.~7 (d) & $1.3\times10^{43}$ (d) & 71 (d) & $1.8\times10^7$ (e)& (f) \\
\hline
NGC 4151 & Fig.~8 (g) & $5\times10^{42}$ (g) & 15.8 (h) & $1.66\times10^{7}$ (i) &   (l) \\
\hline
CGCG 420-015 & (a) & $1\times10^{44}$ (m) & 134 (n) & $5.49\times10^8$ (m)& (o) \\
\hline
\end{tabular}
\caption{Benchmark properties of each galaxy used in our work: we collect the adopted spectral energy distributions (SEDs) for the disk and corona components, the X-ray luminosity integrated between 2~keV and 10~keV, the distance from Earth, the assumed black hole mass, and the source of the Fermi-LAT measurements that we use. Adopted references include (a) \cite{Marconi_2004_SED} (we use this SED parameterization for those galaxies for which an independent measurement is not available), (b) \cite{Padovani:2024ibi}, (c) \cite{Ajello_2023}, (d) \cite{Prince:2025aou}, (e) \cite{2021MNRAS.504.4123N}, (f) \cite{Yang:2025lmb}, 
(g) \cite{Kumar_2024},
(h) \cite{2022ApJ...934..168B},
(i) \cite{2020ApJ...902...26Y},    (l) \cite{Fermi-LAT:2022byn}, (m) \cite{10.1093/mnras/stz3608}, (n) \cite{Tanimoto2018}, (o) \cite{Fermi_sensitivy}.}
\label{tab:astrophysical_params}
\end{table*}

The energy budget of the corona has been addressed in multiple works on the turbulent scenario, and different authors have used a different terminology and focused on different energy scales. Here we provide a concise overview of all the energetics that are dynamically relevant in the corona.

In terms of energy density, the rest-mass energy density, as we have already discussed, is dominated by protons and can be written as $U_{\rm rest}=c^2 m_p \tau_T/R\sigma_T$. Assuming these protons to possess the virial temperature $k_B T_p/m_p c^2\simeq  r_g/3R$, the pressure associated with the thermal protons is
\begin{equation}
    P_{\rm ther}=n_p k_B T_p\simeq \frac{ m_p c^2 r_g \tau_T}{3 \sigma_T R^2}.
\end{equation}
As we will see, the magnetic pressure gradient can be significantly larger than the gravitational energy density, so thermal particles energized by the turbulence may reach substantially higher temperatures. Accordingly, this estimate should more reasonably be associated with the gravitational energy density of the corona, while the protons themselves could be hotter.

The magnetic field possesses, in turn, an energy density $U_B=\sigma U_{\rm rest}/2$ and a turbulent energy density $U_{B, \mathrm{tur}}=\sigma_{\rm tur} U_{\rm rest}/2$. In general, the corona is dominated by magnetic, rather than thermal, pressure. The ratio of thermal to magnetic pressure is given by the beta parameter defined in Eq.~\ref{eq:beta_parameter}.
Since we will find that, for the neutrino excesses reported by IceCube, we usually have $\sigma_{\rm tur}\sim 0.1-1$, with $\sigma=\sigma_{\rm tur}/\eta_B$ potentially even larger, and $R/r_g\gtrsim 10$---otherwise the corona would be so compact that reconnection would be a more plausible acceleration mechanism~\citep{Fiorillo:2023dts,Karavola:2024uui}---we generally have $P_{\rm ther}/U_B\lesssim 1$. A dominance of magnetic over gravitational energy density naturally suggests a highly dynamical environment, which is generally expected also on the ground of high variability across different scales~\citep{Zhao:2025qmi}. We should stress that our estimate of $P_{\rm ther}$ relies on an approximate hydrostatic balance between pressure and gravity; since $U_B$ is however larger than gravitational energy density, one may expect that magnetic dissipation could heat the particles to an approximate equipartition $P_{\rm ther}\sim U_B$. Generally, the imbalance between turbulent magnetic energy and gravitational energy density is suggestive of a non-steady environment; we do not enter here a discussion of the potential dynamics, which would require a better understanding of the coronal geometry and pressure support~\citep{Sridhar:2024rii}. Rather, we use neutrino and gamma-ray measurements to directly infer the energy scales, which should therefore be understood as average values of an intrinsically dynamical corona.

If the corona is to be magnetically powered, the magnetic luminosity $L_B$ needs to be sufficient to power $L_X$, the coronal intrinsic X-ray luminosity over the entire X-ray energy range. We therefore expect $L_X/L_B\lesssim1$, or
\begin{equation}
    \frac{L_X}{L_B}=\frac{3\eta \sigma_T L_X}{2\pi m_p c^3\eta_{\rm rec}\sigma_{\rm tur}^{3/2}\tau_T R}\lesssim1.
\end{equation}
Hence, this sets an upper bound on the maximum X-ray luminosity that the corona can have within a turbulent scenario
\begin{equation}
    \frac{L_X}{R}\lesssim 7.1\times 10^{30}\;\frac{\sigma_{\rm tur}^{3/2}}{\eta}\;\frac{\mathrm{erg}}{\mathrm{s}\,\mathrm{cm}}.
\end{equation}
In turn, the more conventional definition of compactness is limited by
\begin{equation}
    \ell_X= \frac{\sigma_T L_X}{4\pi R m_e c^3} \lesssim 15.3\frac{\sigma_{\rm tur}^{3/2}}{\eta}\,.
\end{equation}
This upper bound strictly descends from the assumption that the corona is magnetically dominated, so it can be evaded if a different mechanism is proposed for X-ray emission, at the cost of more free parameters.

Finally, the non-thermal protons introduce a new energy budget $L_p$; by definition $L_p/L_B=\mathcal{F}_p\lesssim1$, since we assume that proton acceleration is powered by turbulence. With this definition, we can now consider the dynamical impact that non-thermal protons have on the coronal turbulence. The ratio of their energy density to the turbulent magnetic energy density is
\begin{equation}\label{eq:energy_density_ratio}
    \frac{U_{p,\rm nth}}{U_{B, \rm tur}}=\frac{5 \mathcal{F}_p \eta_{\rm rec}}{2\sqrt{\sigma_{\rm tur}}}.
\end{equation}
In general, we find that the ratio of these energy densities is smaller than $\mathcal{F}_p$, which itself is smaller than unity, unless very small values of $\sigma_{\rm tur}$ are assumed, which we find to be disfavored. This implies that non-thermal protons remain energetically subdominant with respect to the magnetized turbulence that powers them. This justifies treating the turbulence as evolving independently of the non-thermal proton population, which does not exert a significant back-reaction (but see \cite{Lemoine:2024wsw}).

Parallel lines of research on the subject~\citep{Murase_2020,Lemoine:2025roa} have discussed the energetics of non-thermal protons primarily by means of their relative pressure contribution in comparison with the thermal one; since the pressure of a relativistic gas is $P_{p,\rm nth}\simeq U_{p,\rm nth}/3$, this ratio is roughly
\begin{equation}
    \frac{U_{p,\rm nth}}{3P_{\rm ther}}\simeq \frac{5 \mathcal{F}_p R \eta_{\rm rec}\sqrt{\sigma_{\rm tur}}}{16 r_g}.
    \label{eq:feedback}
\end{equation}
As argued above, the pressure in the corona is largely magnetically dominated, and a more appropriate measure for a possible dynamical feedback of non-thermal protons on the coronal evolution is given by the ratio $P_{p,\rm nth}/U_{B,\rm tur}$, which is one third of Eq.~\ref{eq:energy_density_ratio}.
It is then clear that with $\mathcal{F}_p <1$, the relative contribution of non-thermal protons to the coronal magnetic pressure is expected to remain subdominant.

Overall, the dominant constraint on the energetics of non-thermal protons in the corona is $L_p\lesssim L_B$, i.e. $\mathcal{F}_p\lesssim 1$. If this constraint is satisfied, the contribution of non-thermal protons to the coronal pressure and energy density turns out to be smaller than the magnetic one.

\section{Methods: analysis of Seyfert galaxies}\label{sec:analysis}

The main aim of this work is to demonstrate the potential of neutrino observations to constrain the internal properties of AGN coronae. For this purpose, we apply our model to the four galaxies for which IceCube reported an excess \citep{Abbasi:2025tas}: NGC 1068, NGC 4151, CGCG 420-15 and NGC 7469. Among these four galaxies, a global excess is reported by IceCube only for NGC~1068 and NGC~7469, with corresponding global significance of 4.0 $\sigma$ for NGC~1068 (for a sample of 110 gamma-ray bright sources) and 2.4 $\sigma$ for NGC~7469 (for a sample of 47 X-ray bright AGN). In the main text, we will consider only these two galaxies, which also happen to be the ones for which we find the clearest constraints on the parameters of the magnetized corona. In Appendix~\ref{appendix:altre}, we also show the results for NGC~4151 and CGCG~420-015, which however are less conclusive. 

In this section, we describe how our method, based on the combination of neutrino observations and Fermi-LAT information, allows to constrain the model parameters of neutrino-emitting AGN coronae. The assumed properties for each of the analyzed galaxies are collected in Table~\ref{tab:astrophysical_params}. 

\subsection{Multimessenger emission}\label{sec:multimessenger_emission}

\begin{figure*}[t]
    \centering
    \includegraphics[width= \textwidth]{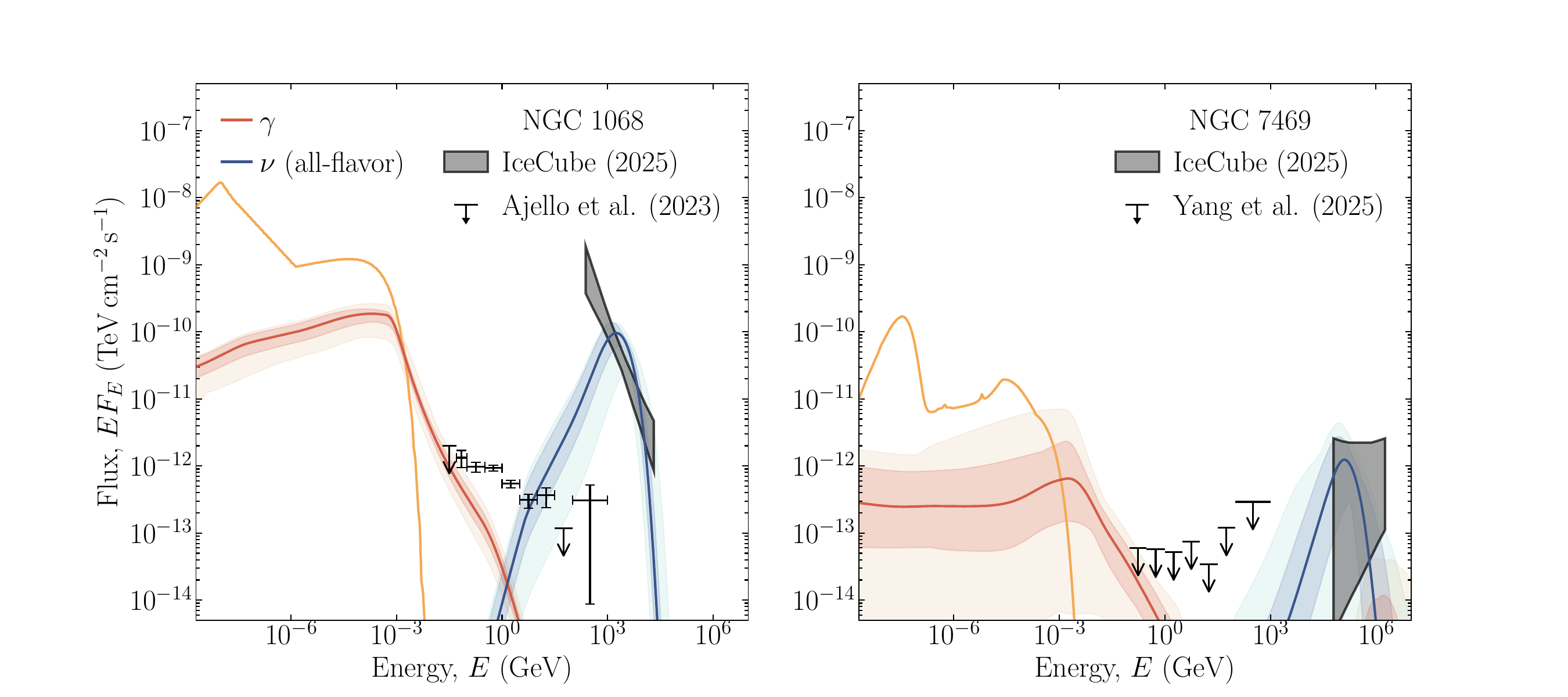}
    \caption{Neutrino and electromagnetic emission from the coronal region, for NGC~1068 (left) and NGC~7469 (right). We show the $68\%$ and $95\%$ confidence level (C.L.) regions with different shadings, together with the predicted best-fit curves. Since the confidence regions are shown mostly as a visual guide, we use the $\chi^2$ threshold for the TS corresponding to one degree of freedom, so as to match the exclusion contours for the individual parameters shown in Figs.~\ref{fig:1068_param} and~\ref{fig:7469_params}. The assumed SED is shown in orange for both galaxies. We do not include extragalactic gamma-ray absorption, which would only affect very-high-energy gamma rays above hundreds of GeV. All Fermi-LAT measurements are used only as upper bounds to leave space for additional contributions from different emission regions.}
    \label{fig:1068}
\end{figure*}

The predicted neutrino signal in our model depends sensitively on the proton cooling rate, in turn affected by the spectral energy distribution (SED) of the galaxy. For those galaxies for which an independent estimate of the broadband SED is available in the literature, we adopt it directly from the corresponding references listed in Table~\ref{tab:astrophysical_params}. When this estimate is not available, we use $L_X^{\rm 2-10\,keV}$, the measured X-ray luminosity between $2-10\,\mathrm{keV}$ corrected for absorption, to obtain the empirical estimate of the SED from the parameterization of~\cite{Marconi_2004_SED}. For these SEDs we assume a cutoff energy of $E_{\mathrm{cut}} = 511\,\mathrm{keV}$, but we stress that this value is not directly observed, with widely varying estimates for different sources. The uncertainty propagates primarily on the absorption rate of $\mathrm{MeV}$ photons, so it does not change our conclusions based on the internal absorption of photons in the Fermi-LAT energy range. On the other hand, the uncertainty in $E_{\mathrm{cut}}$ limits the accuracy with which one can derive the total intrinsic X-ray luminosity $L_X$ from the measured luminosity in the $2-10\,\rm keV$ band.
For this reason, we adopt $L_X^{\rm 2-10\,keV}$ as a proxy of the total power output in X-rays, $L_X$, and use the parameter
\begin{equation}\label{eq:FX}
\mathcal{F}_X \equiv \frac{L_X^{\rm 2-10\,keV}}{L_B}
\end{equation}
to compare the coronal luminosity the X-ray 2-10 $\rm keV$ band and the magnetic field dissipation rate $L_B$.
In Table~\ref{tab:astrophysical_params} we list for each source the X-ray luminosity in the $2-10\,\rm keV$ band, the black hole mass and the distance assumed in this work.

\begin{figure*}[t!]
    \includegraphics[width=\textwidth]{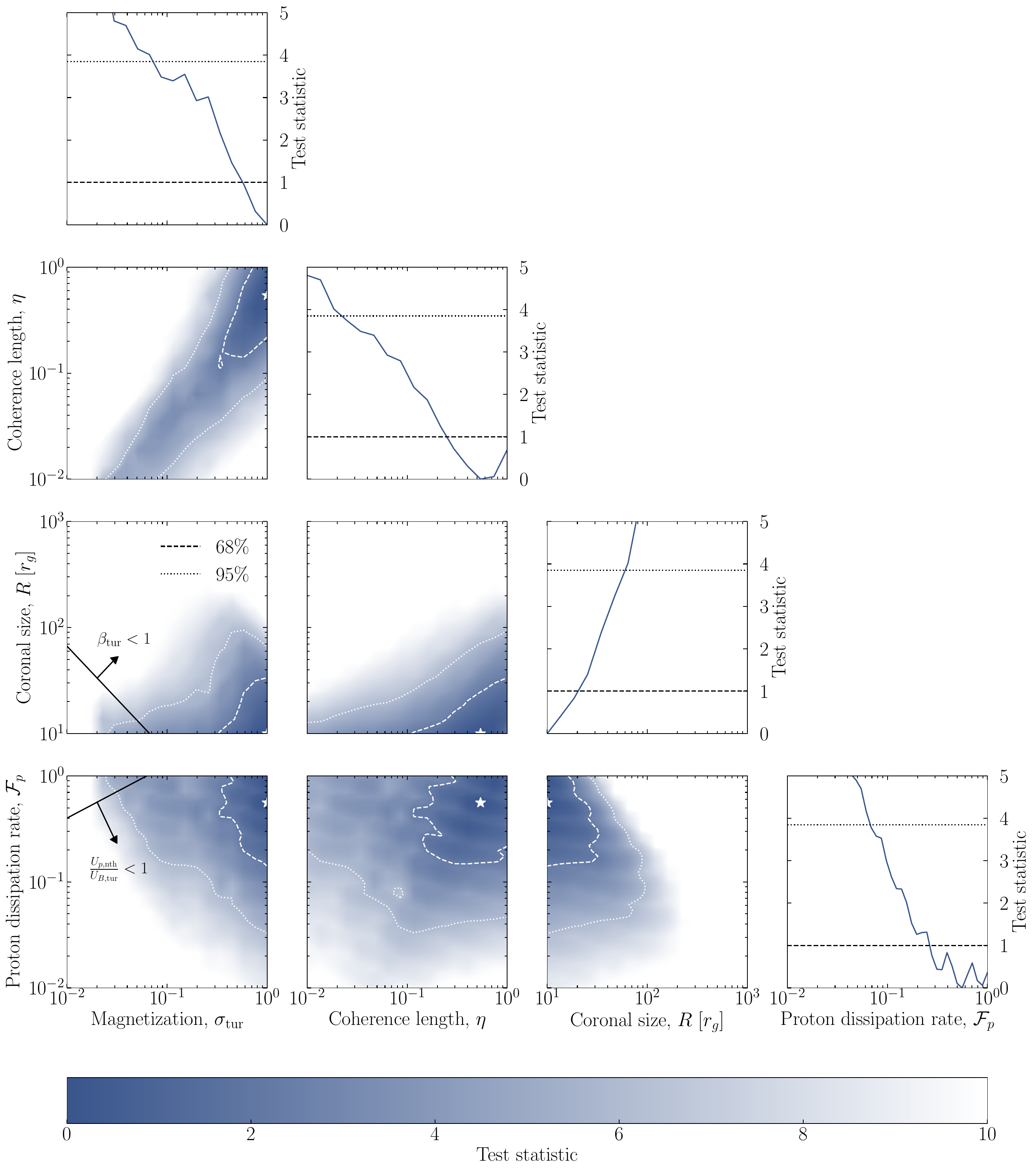}
    \caption{Constraints on coronal properties of NGC~1068 from the combined neutrino and gamma-ray signal. We highlight the regions of parameter space where $\beta_{\rm tur}<1$, signaling a strongly magnetized plasma, and $U_{p,\rm nth}/U_{B,\rm tur}<1$, signaling a negligible dynamical feedback of non-thermal protons on the turbulence.}\label{fig:1068_param}
    
\end{figure*}

\begin{figure*}[t!]
    \includegraphics[width=\textwidth]{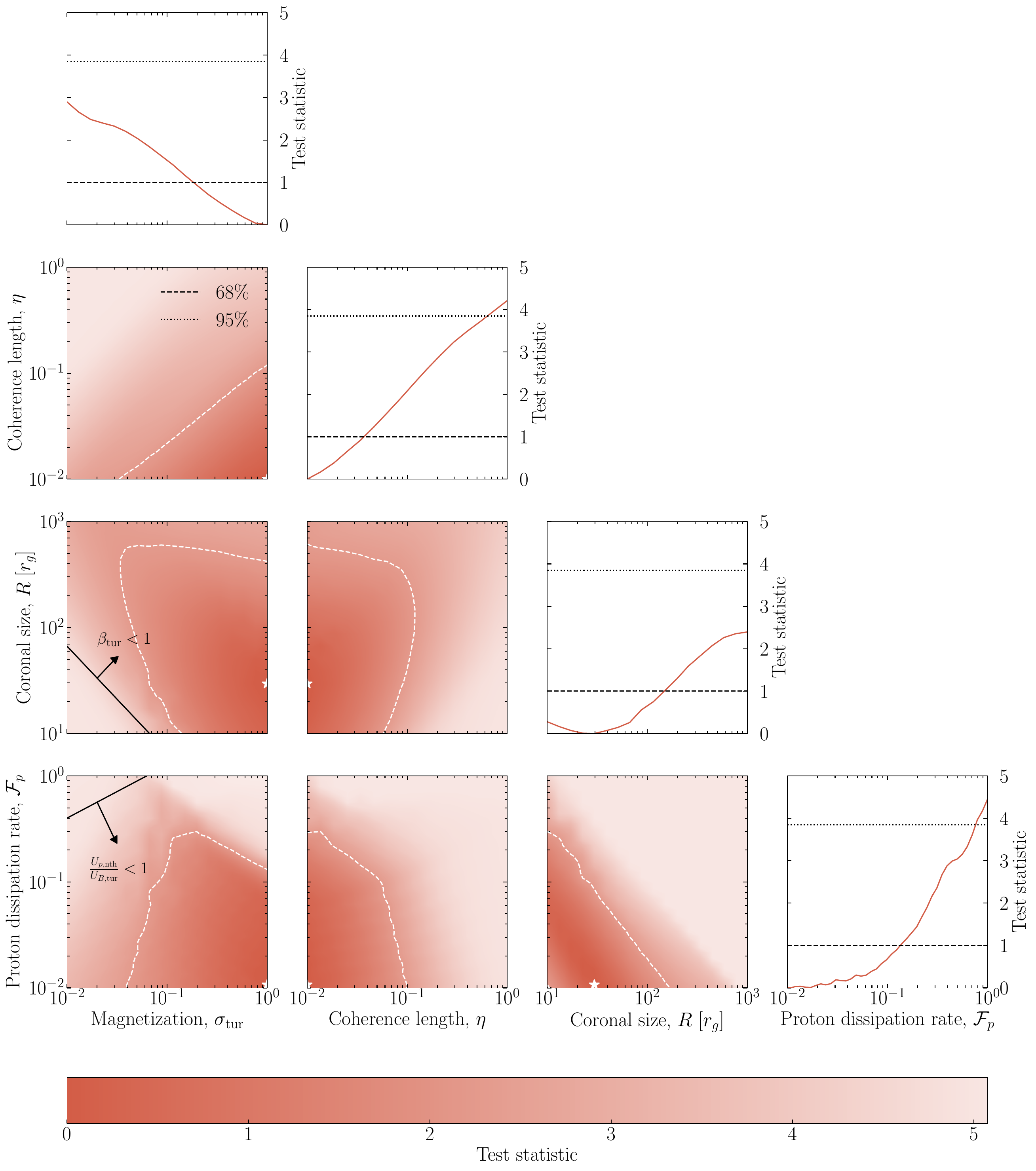}
    \caption{Constraints on coronal properties of NGC~7469 from the combined neutrino and gamma-ray signal. We highlight the regions of parameter space where $\beta_{\rm tur}<1$, signaling a strongly magnetized plasma, and $U_{p,\rm nth}/U_{B,\rm tur}<1$, signaling a negligible dynamical feedback of non-thermal protons on the turbulence.}\label{fig:7469_params}
    
\end{figure*}

\begin{table*}[t]
\centering
\begin{adjustbox}{max width=\textwidth}
\begin{tabular}{|l|c|c|c|c|c|c|c|}
\hline
Galaxy & $\sigma_{\rm tur}$ & $\eta$& $R/r_g$ & $\mathcal{F}_p$ & $\mathcal{F}_X$ & $\beta$ & $U_{p,\rm nth}/U_{B,\, \rm tur}$\\
\hline
NGC 1068 & 1 & 0.5 & 10 & 0.5 & 0.2 & 0.07 & 0.125\\
\hline
NGC 7469 & 1 &  0.01 & 30 & 0.01 & $2\times10^{-4}$ & 0.02 & $3\times 10^{-3}$\\
\hline
NGC 4151 &   0.8 & 1 & 10 & 0.15 & 0.04 & 0.08 & 0.04 \\
\hline
CGCG 420-015 & 0.6   & 1 & 13 & 0.14 & 0.04 & 0.08 & 0.04\\
\hline
\end{tabular}
\end{adjustbox}
\caption{Best-fit values for the model parameters $\sigma_{\rm tur}$, $\eta$, $R/r_g$, $\mathcal{F}_p$ for all of the galaxies considered. We also report the derived parameters $\mathcal{F}_X$, $\beta$, and $U_{p,\rm nth}/U_{B,\rm tur}$.}
\label{tab:best-fit}
\end{table*}

For a given galaxy, the resulting proton distribution and the electromagnetic and neutrino signal depend on the internal parameters of the corona, namely the magnetization $\sigma_{\rm tur}$, the dimensionless coherence length $\eta$, the coronal radius $R$, and the fraction of magnetically dissipated energy transferred to non-thermal protons $F_p$. Another important input is the target photon density: we obtain it as $n_{\rm ph}=3L_{\rm ph}/(4\pi R^2 c)$, where $L_{\rm ph}$ is inferred from the adopted SED. We note here a factor of three difference compared to earlier works~\citep{Murase:2019vdl, Murase_2022ApJL, Fiorillo:2024akm,Fiorillo:2025ehn,
Yang:2025lmb} arising purely from geometric considerations, since we assume here the photons to be uniformly injected in the spherical region. None of these prescriptions is anyway strictly realistic, since the dominant target radiation for Bethe-Heitler interactions is produced in outer regions of the accretion flow which are anisotropic, so this discrepancy is rooted in the unavoidable uncertainty on the coronal geometry.

We track the dependence of neutrino and cascaded electromagnetic emission on these parameters by explicitly solving the transport equation for protons, Eq.~\ref{eq:transport}, using the same method as in~\cite{Fiorillo:2024akm}.
We then use the resulting steady-state proton distribution as input for the code $\texttt{AM}^3$ \citep{Klinger:2023zzv}, which solves the transport equation for the particle species present in the radiation zone. By injecting in the radiation zone (here identified with the corona) the steady-state proton distribution and the target photon field $n_{\rm ph}$ over the dynamical time $R/c$, we compute numerically the steady-state emission of photons and neutrinos from the corona. At this second stage we neglect the cooling and acceleration of protons by switching off the relative terms in the $\texttt{AM}^3$ solver, since it was already included in the first step to determine the steady-state proton distribution. Although this two-step calculation neglects potential interaction of protons with photons of secondary emission, \cite{Yuan:2025ctq} showed that even when these interactions are taken into account in a time-dependent solution (i.e. solving iteratively for the accelerated protons and the radiation, so that also secondary radiation is included as target in the cooling terms), they are irrelevant for this coronal environment, and the dominant photon target for the interactions of all particles is the ambient photon field of the AGN $n_{\rm ph}$.

The neutrino and electromagnetic emission depend in a complex way on the input parameters, due to the interplay between the acceleration efficiency, well captured by our parametric estimates, and the cooling processes, sensitively dependent on the assumed SED and difficult to capture parametrically. Although the spectral shape of electromagnetic cascades in strongly magnetized environments can exhibit universal properties \citep{Fiorillo:2025kuh}, in these coronae the abundant production of secondary pairs through Bethe-Heitler processes  hinders a parametric approach. We therefore base our results on the full numerical calculations, which we also make publicly available on GitHub~\citep{GitRepo}. For these calculations we used a logarithmically spaced grid of 18 points for $\sigma_{\rm tur}$ in the range [0.01, 1], 16 points for $\eta$ in the range [0.01, 1] and 18 points for $R/r_g$ in the range [10, 1000].

\subsection{Statistical analysis}\label{sec:statistical_analysis}
The next step is to compare the photon and neutrino spectra emitted by a corona described by the model parameters $\boldsymbol{\theta}\equiv$ ($\sigma_{\text{tur}},\,\eta,\,R,\,\mathcal{F}_p$) with the observations.
For neutrinos, our goal is to fit the excesses recently reported by the IceCube Collaboration \citep{Abbasi:2025tas}. The data and the effective areas used in that analysis are not public, so we rely on the publicly available code $\texttt{PLE} \nu \texttt{M}$ \citep{Schumacher:2025qca} to retrieve an approximate dataset from the published results. $\texttt{PLE} \nu \texttt{M}$ simulates the response of the IceCube detector to an input flux of neutrinos and computes the two-dimensional histogram of events in bins of reconstructed energy of the neutrino $E^{\text{reco}}$ and of $\psi^2$, where $\psi$ is the angular distance between the reconstructed direction of the detected neutrino and the source location. Using an exposure time of 13.1 years, for each galaxy we set the declination of the source and the input fluxes to the best-fit reported by IceCube \citep[][]{Abbasi:2025tas, Bellenghi:2024dcp}. With this setup we compute in $\texttt{PLE} \nu \texttt{M}$ the expected number of astrophysical track events $a_{ij}$ and background events $b_{ij}$, both distributed in $E_i^{\text{reco}}$ and $\psi^2_j$. From these we construct the \textit{mock} dataset $k_{ij} = a_{ij} + b_{ij}$, that we use in lieu of the true events detected and analyzed by IceCube.

For a corona with model parameters $\boldsymbol{\theta}$, we use its flux of $\nu_\mu + \overline{\nu}_\mu$, combined with $\texttt{PLE} \nu \texttt{M}$, to obtain the resulting \textit{test} astrophysical events $\alpha_{ij}$. For the flux of $\nu_\mu + \overline{\nu}_\mu$ at Earth, we take 1/3 of the all-flavor flux emitted by the corona and computed in our model. Indeed neutrinos are mostly emitted in the pion-beam regime, where the effect of synchrotron cooling on the parent muons and pions is negligible and does not alter the flavor composition at production. Thus, in this scenario, neutrinos are produced with a flavor ratio ($\nu_e\,$:$\,\nu_\mu\,$:$\,\nu_\tau$) = (1:2:0), and after oscillations they arrive at Earth with composition (1:1:1), so that for each flavor the flux of neutrinos and antineutrinos is roughly 1/3 of the total flux.\footnote{This should be contrasted with the alternative coronal reconnection scenarios, where the more compact region leads to stronger magnetic fields and a more significant impact of muon cooling~\citep{Karavola:2026rpg}.}\\ In analyzing the mock data, the spectral shape of the atmospheric background at the source declination is fixed, and obtained by the $\texttt{MCEq}$ library \citep{Fedynitch:2015zma} used in \plenum, but we allow for a rescaling of its normalization up to $\pm 50\%$ through a nuisance parameter $\mathcal{B}$, so the predicted event distribution is $\mu_{ij}(\boldsymbol{\theta}, \mathcal{B}) = \alpha_{ij}(\boldsymbol{\theta}) + \mathcal{B}b_{ij}$. From the mock data set $k_{ij}$, and the theoretical predicted distribution $\mu_{ij}(\boldsymbol{\theta},\mathcal{B})$, we construct a two-dimensional multi-Poissonian likelihood $\mathcal{L}_\nu(\boldsymbol{\theta},\mathcal{B})$ including all bins in energy and angular distance. The associated chi-squared, defined as $\chi^2_\nu=-2\ln\mathcal{L}_\nu$, can be written
\begin{equation}
    \chi^2_\nu(\boldsymbol{\theta}, \mathcal{B}) =  2\sum_{E^{\mathrm{reco}}_i,\psi^2_j }\left[\mu_{ij}(\boldsymbol{\theta}, \mathcal{B}) - k_{ij}\ln(\mu_{ij}(\boldsymbol{\theta}, \mathcal{B}))\right].
\end{equation}
We do not include the term $\ln(k_{ij}!)$, which is independent of $\boldsymbol{\theta}$ and therefore drops out from the minimization we perform later.

In addition to neutrinos, we consider also gamma-ray observations, using the measurements and upper bounds from Fermi-LAT to further constrain the coronal properties. Gamma rays may also be produced from other regions beyond the corona: for instance, in the NGC~1068 case, these include the weak jet~\citep{Lenain2010}, the circumnuclear starburst region~\citep{Yoast-Hull2013,Ambrosone:2021aaw,Eichmann_2022}, a large scale AGN-driven outflow~\citep{Lamastra_2016}, failed line-driven winds~\citep{Inoue_S_2022}, or an ultra-fast outflow~\citep{Peretti_2023}. Since it is unclear whether the observed gamma-ray emission is truly coronal, here we simply require the coronal emission not to exceed the data, treating all Fermi-LAT observations as upper limits and allowing for additional contributions from other emitting regions. We use the available analysis in the literature for NGC 1068 \citep{Ajello_2023} and NGC 7469 \citep{Yang:2025lmb}. 
Similarly to what we did for neutrinos, from the model photon flux we determine for each energy bin $n$ the integrated flux level $\phi_{\gamma, \, n}(\boldsymbol{\theta})$. By assuming 0 events and treating the Fermi-LAT flux point ${\Phi_n}$ as a $2\sigma$ upper limit, we obtain the chi-squared as
\begin{equation}
    \chi_\gamma^2(\boldsymbol{\theta}) = \sum_n \frac{ \phi_{\gamma, \, n}(\boldsymbol{\theta})^2}{(\Phi_n/2)^2}\,.
\end{equation}
Finally, we sum the two terms and minimize with respect to the nuisance parameter $\mathcal{B}$ to obtain the test statistic (TS)
\begin{equation}
    \Upsilon (\boldsymbol{\theta}) = \min_\mathcal{B}\left[\chi^2_\nu(\boldsymbol{\theta}, \mathcal{B}) + \chi_\gamma^2(\boldsymbol{\theta})\right] \,.
\end{equation}
In the asymptotic limit of many detected events of ~\citep{Wilks:1938dza}, the test statistic (TS) follows a $\chi^2$ distribution with 4 degrees of freedom, which allows us to construct the allowed regions in the four-dimensional parameter space. Nevertheless, for visual clarity, we will always show confidence regions for the marginalized TS (in the frequentist sense, i.e. minimized with respect to the complementary parameters) for the pairs of variables $\sigma_{\rm tur}-\eta$ and $\mathcal{F}_p-R$, as well as the marginalized TS for each of the four parameters individually. These marginalized TSs should be distributed as $\chi^2$ variables with 2 and 1 degrees of freedom respectively.

\section{Results: combined fit of coronal properties}\label{sec:fit_coronal}

The multimessenger emission predicted by our analysis is shown in Fig.~\ref{fig:1068}, directly compared with the measurements. For neutrinos, the most natural comparison is with the uncertainty band reported by the IceCube collaboration for a power-law fit~\citep{Abbasi:2025tas}. The IceCube collaboration has also performed~\citep{IceCube24_Seyfert} statistical fits of the observations based on a disk-corona model from~\cite{Murase_2020,Kheirandish_2021}, which however was taken with a fixed spectral shape. In our fit procedure, the variation of the four parameters $\boldsymbol{\theta}$ allows for considerable variation in the spectral shape. In particular, the position of the neutrino peak can vary by orders of magnitude, with larger $\sigma_{\rm tur}$ and $R$, and smaller $\eta$, favoring higher neutrino peak energies.

Our analysis for NGC~1068 favors a neutrino flux peaking at $\sim1$~TeV. In this sense, the preferred solution differs from our previous approach in~\cite{Fiorillo:2025ehn}, where we qualitatively matched our model predictions to the reported power-law fit in~\cite{IceCube:2022der}, leading to neutrinos peaking at $\sim10$~TeV. The preferred neutrino flux at the peak is also slightly larger, reaching around $E F_E\simeq 10^{-10}\,\mathrm{TeV}\,\mathrm{cm}^{-2}\,\mathrm{s}^{-1}$. The gamma-ray measurements contribute to the fit primarily by requiring a compact enough region, so that the cascaded photon flux, shown in red in Fig.~\ref{fig:1068}, has a sharp cutoff at $\sim1$~MeV due to $\gamma\gamma$ absorption, and is therefore consistent with the upper bounds from Fermi-LAT.

For NGC~7469, the neutrino flux peaks at significantly higher energies, at $\sim100$~TeV, which is consistent with the reported detection of two events in this energy range. The best-fit multimessenger solution involves again a cutoff in the gamma-ray flux at a few MeV.

In terms of coronal properties favored by the analysis, Table~\ref{tab:best-fit} collects the best-fit values for the model parameters for NGC~1068 and NGC~7469, as well as for the other two galaxies from which IceCube has reported an excess~\citep{IceCube24_Seyfert} which are discussed in more detail in Appendix~\ref{appendix:altre}. We show the complete results of our study for NGC~1068 and NGC~7469 in Fig.~\ref{fig:1068_param} and~\ref{fig:7469_params}, where we report a corner plot showing the likelihood and confidence regions across the entire parameter space.

For NGC~1068, the measurements identify a clearly defined region in the parameter space, with a significant portion of it excluded at 95\% confidence level. The best-fit explanation requires a compact corona ($R\simeq 10\,r_g$), strongly magnetized ($\sigma_{\rm tur}\simeq 1$), with a rather large coherence length ($\eta\simeq 0.5$) and a large fraction of energy dissipated in protons ($\mathcal{F}_p\simeq 0.5$). This broadly agrees with the expectations outlined in~\cite{Fiorillo:2024akm,Fiorillo:2025ehn} based on a qualitative matching to the IceCube power-law fit, although we now favor a more compact corona and a larger value for $\eta$, lowering the typical neutrino energy (see the discussion in Sec.~\ref{sec:back_of_the_envelope}). The gamma-ray upper bounds also lead to a strong constraint on the coronal radius, which must be sufficiently small to provide the required gamma-ray opacity, an argument discussed in model-independent terms in~\cite{Murase_2022ApJL}. The combined multimessenger analysis shows a clear degeneracy among $\sigma_{\rm tur}$, $\eta$, and $R$, mostly due to the peak neutrino energy, which depends on the combination of parameters $\sigma_{\rm tur} R/\eta$ (see our discussion in Sec.~\ref{sec:back_of_the_envelope}), implying a positive degeneracy between $\sigma_{\rm tur}$ and $\eta$. The radius $R$ is so strongly constrained that its degeneracy with the other parameters is weaker, although still visible. Overall, we find a clear lower bound on the magnetization, $\sigma_{\rm tur}\gtrsim 0.02$ at the 95\% confidence level.

In Fig.~\ref{fig:1068_param} we also highlight the region where $\beta_{\rm tur}=\beta/\eta_B<1$, which encompasses essentially the entire allowed parameter space and therefore favors a strongly magnetized plasma. We further mark the region where $U_{p,\rm nth}/U_{B, \rm tur}<1$, namely the region where non-thermal protons have negligible dynamical feedback on the turbulence. This condition is also fulfilled across most of the allowed parameter space, and certainly for our best-fit solutions.

The best-fit solution from our model lies at the edge of the allowed parameter space, preferring a corona as compact and magnetized as possible. This shows that the theoretical priors truly play a key role in deciding the favored explanation: a more magnetized and more compact corona ($\sigma_{\rm tur}\gg 1$, $R\lesssim 10\,r_g$) would more reasonably be associated with the magnetospheric region, where dissipation likely happens by means of strongly magnetized reconnection layers~\citep{Fiorillo:2023dts,Karavola:2024uui, Karavola:2026rpg}. More generally, we cannot extend our models at lower radii without accounting for geometrical suppression of the optical/UV disk emission, which is likely produced farther out.

For NGC~7469, the more limited statistics leads to much less defined exclusion regions. Nevertheless, at 68\% confidence level, there is a favored region centered on a best-fit solution that involves a strongly magnetized ($\sigma_{\rm tur}\simeq 1$), compact ($R\simeq 30\,r_g$) corona. In this case, the preferred coherence length is very small ($\eta\simeq 0.01$), which maximizes the acceleration rate of non-thermal protons as needed to reproduce the observed neutrino emission at higher energies. The high energy of the non-thermal protons, necessary to explain the ~100 TeV neutrinos, must remain consistent with a relatively small neutrino flux, which in turn requires a small fraction of dissipated energy transferred to non-thermal protons ($\mathcal{F}_p\simeq 0.01$). At the 95\% confidence level, the two-dimensional parameter spaces are basically unconstrained within the parameter ranges explored. For the individual parameters, using the one-parameter marginalized likelihood, we find an upper bound at 95\% confidence level on $\mathcal{F}_p$, signaling that too high a proton efficiency would overproduce the neutrino signal and therefore would be excluded. On the other hand, we do not find a \textit{lower} bound on $\mathcal{F}_p$, which implies that, with our statistical methods, the mock data are still compatible with a pure background model. This is likely due to our use of a mock data sample, rather than the real IceCube data which are at present not public. Therefore, an analysis of the actual IceCube observations may likely reveal much more stringent and reliable constraints on the model.

For the best-fit solution of each source, the ratio between the steady state density of non-thermal protons $n_{p, \rm nth}$ and the density of thermal protons in the corona $n_{p, \rm th}$ obtained assuming charge neutrality is generally very small, $\mathcal{O}(10^{-6}\div10^{-8})$, similarly to what we found in~\cite{Fiorillo:2024akm}. This ratio is highly model-dependent, and in principle this issue might be relieved if the proton spectra were somewhat softer, as discussed in more detail in~\cite{Fiorillo:2025ehn}: for example, an additional escape channel with an energy-independent timescale $t_{\rm esc}$, such as the hydrodynamical escape timescale proposed in~\cite{Lemoine:2025roa}, would soften the proton spectrum to $dN_p/dE_p\propto E_p^{-1-t_{\rm acc}/t_{\rm esc}}$. Hence, there would be more protons injected in the low-energy range, although at the cost of a somewhat higher required proton luminosity $L_p$. While we do not investigate this explanation in more detail here -- we showed the impact of hydrodynamical escape in~\cite{Fiorillo:2025ehn} -- we notice that it would not strongly affect the region close to the peak of the neutrino spectrum, since in this energy region the cooling timescale becomes much shorter than both the acceleration and the escape timescale. Similar considerations hold for the ratio between the number of non-thermal secondary pairs produced by the accelerated protons $n_{e^{\pm}, \rm nth}$ and the number of thermal pairs estimated by the Compton thickness argument in Eq.~\ref{eq:pair_density}. These secondary pairs are produced via Bethe-Heitler pair production, as the decay products of photopion interactions, and from $\gamma \gamma$ pair production of photons on the target SED of the accretion flow. The value of this ratio for pairs is $\lesssim 10^{-3}$, meaning that the pairs stemming from the interaction of accelerated protons have a minor impact on the Comptonization of the disk photons in the corona, which is still predominantly determined by the primary electrons.


\section{Back-of-the-envelope neutrino constraints on coronal properties}\label{sec:back_of_the_envelope}

Our statistical analysis, which goes into as much detail as public IceCube data allow, has revealed several stringent constraints from neutrino observations on the coronal properties. In this section, we now show how these constraints can be understood on the basis of rather simple estimates. As in the previous sections, our primary parameters are $\sigma_{\rm tur}$ (we assume for simplicity $\eta_B=1$, although $\eta_B$ is mostly degenerate with $\sigma$, the total magnetization), $\eta$, $\mathcal{F}_p$, and $R_{10}=R/10 r_g$. 

\subsection{NGC~1068}

For the case of NGC~1068, the acceleration timescale from Eq.~\ref{eq:acceleration_rate} gives
\begin{equation}
    t_{\rm acc}\simeq 3\times 10^3\,\frac{\eta R_{10}}{\sigma_{\rm tur}}\,\mathrm{s}.
\end{equation}
The neutrino signal inferred by IceCube drops above a few TeV; since in photohadronic interactions neutrinos from meson decays are produced with a typical energy of around $5\%$ of the parent proton energy, this implies that the energy of the parent protons is around tens of TeV. With the SED we have assumed, at a proton energy of $E_{p,\rm max}=20\,\mathrm{TeV}$, energy losses are mildly dominated by $p\gamma$ interactions  --- an interesting difference from earlier works~\citep{Murase_2020,Fiorillo:2024akm} in which Bethe-Heitler dominates over $p\gamma$, due to the more accurate determination of $p\gamma$ energy loss rate included in \texttt{AM}$^3$ --- with a timescale
\begin{equation}
    t_{p\gamma}\simeq 900\, R_{10}^2\,\mathrm{s} .
\end{equation}
The quadratic dependence on the coronal radius reflects the geometric dilution of the photon number density in a larger environment. For comparison, the timescale for Bethe-Heitler losses is larger by a factor of two. Hence, assuming that photohadronic cooling balances proton acceleration at around $E_{p,\rm max}$, we obtain a constraint
\begin{equation}\label{eq:first_constraint_1068}
    \frac{\sigma_{\rm tur}R_{10}}{\eta}\sim3.
\end{equation}
Since $p\gamma$ interactions dominate, the majority of proton energy is converted to the products of $p\gamma$ scattering, which include gamma rays, neutrinos, and $e^\pm$. The fraction of energy channeled in neutrinos, assuming for example only the contribution from $\Delta$ resonance, is $1/4$~(e.g.~\cite{Winter:2012xq,Fiorillo:2024jqz}). On the other hand, the competition of other interaction channels such as Bethe-Heitler interactions, as well as the presence of a significant amount of energy passed to lower energy protons, may significantly reduce this fraction. Hence, we can benchmark the ratio of neutrino-to-proton-luminosity $L_\nu/L_p$ directly from our numerical treatment: for the best-fit solution for NGC~1068, shown in Fig.~\ref{fig:1068}, we find $L_\nu/L_p\simeq0.05$. The reported IceCube signal provides us with a rough estimate of the neutrino luminosity: with a flux $\int dE F_E\sim 2\times 10^{-10}\,\mathrm{TeV}\,\mathrm{cm}^{-2}\,\mathrm{s}^{-1}$, we deduce a neutrino luminosity $L_\nu\sim \int dEF_E \times 4\pi d_L^2\sim 4\times 10^{42}\,\mathrm{erg/s}$. Since the magnetic dissipation rate is~(e.g. Eq.~(17) of~\cite{Fiorillo:2025ehn})
\begin{equation}
    L_B\simeq 10^{44}\, \frac{\sigma_{\rm tur}^{3/2}R_{10}}{\eta} M_7\; \mathrm{erg/s},
\end{equation}
where $M_7$ is the black hole mass in units of $10^7\,M_{\odot}$, we find
\begin{equation}
    L_p\simeq 6.7\times 10^{43}\,\frac{\mathcal{F}_p\sigma_{\rm tur}^{3/2}R_{10}}{\eta}\, \mathrm{erg/s}.
\end{equation}
Using the constraint from Eq.~\ref{eq:first_constraint_1068}, and matching to the rough estimate for the neutrino luminosity obtained above, we find a new constraint
\begin{equation}\label{eq:second_constraint_1068}
    \mathcal{F}_p \sigma_{\rm tur}^{1/2}\sim 0.4.
\end{equation}
Finally, we may use Eq.~\ref{eq:FX} to obtain
\begin{equation}
    \mathcal{F}_X\simeq 0.43\frac{\eta}{\sigma_{\rm tur}^{3/2}R_{10}}\sim\frac{0.1}{\sigma_{\rm tur}^{1/2}},
\end{equation}
which shows that the corona is generally consistent with being magnetically powered. Since $L_X\sim 0.1 L_{\rm bol}$, where $L_{\rm bol}$ is the bolometric AGN luminosity, the magnetic dissipation rate is comparable with the bolometric luminosity. Since the latter is usually assumed to be a fraction of the accretion power, the magnetic dissipation rate is consistently smaller than the accretion power. 

Our back-of-the-envelope scalings also justify the general degeneracies found among different parameters in the analysis of Fig.~\ref{fig:1068_param}, in particular the positive correlation among $\sigma_{\rm tur}$ and $\eta$, since their ratio is what fixes the peak neutrino energy according to Eq.~\ref{eq:first_constraint_1068}.

Finally, given the constraints we have derived, we can apply the general energy-budget considerations discussed in Sec.~\ref{sec:energetics} to deduce the relative census of thermal and non-thermal protons compared to the magnetic turbulence. In particular, the ratio between the energy non-thermal proton energy density and turbulent magnetic energy density is
\begin{equation}
    \frac{U_{p,\rm nth}}{U_{B,\rm tur}}=\frac{5\mathcal{F}_p\eta_{\rm rec}}{2 \sqrt{\sigma_{\rm tur}}}\simeq \frac{0.1}{\sigma_{\rm tur}},
\end{equation}
where we have used Eq.~\ref{eq:second_constraint_1068}. This points once more, as already stressed in~\cite{Fiorillo:2024akm,Fiorillo:2025ehn}, to a small impact of non-thermal protons on the turbulent cascade itself. The turbulent beta parameter is $\beta_{\rm tur}=0.07 (R_{10}\sigma_{\rm tur})^{-1}$, so there are also strong arguments pointing to $\beta_{\rm tur}\lesssim 1$.

\subsection{NGC~7469}

We can use similar arguments to determine how these constraints change for NGC~7469. In this case, the acceleration timescale changes only by virtue of the different black hole gravitational radius
\begin{equation}
    t_{\rm acc}\simeq 9\times 10^3\,\frac{\eta R_{10}}{\sigma_{\rm tur}}\,\mathrm{s}.
\end{equation}
The neutrino signal inferred by IceCube reaches a few hundreds of TeV; by the same argument used for NGC~1068, this implies protons being accelerated up to energies of $E_{p,\rm max}\simeq 2\,\mathrm{PeV}$. At this energy, with the SED that we assume for NGC~7469, we find that energy losses are dominated by $p\gamma$ interactions, with a typical timescale
\begin{equation}
    t_{p\gamma}\simeq 50\,R_{10}^2\,\mathrm{s}.
\end{equation}
To achieve proton acceleration up to $\sim 2\,\rm PeV$, the constraint of Eq.~\ref{eq:first_constraint_1068} on $\sigma_{\rm tur}$, $R$, and $R_{10}$ now becomes
\begin{equation}\label{eq:first_constraint_7469}
    \frac{\sigma_{\rm tur}R_{10}}{\eta}\sim 180. 
\end{equation}
Following the same steps as for NGC~1068, we now turn to the normalization of the neutrino spectrum. The peak flux reported by IceCube reaches $\int dE F_E \sim 2\times 10^{-12}\,\mathrm{TeV}\,\mathrm{cm}^{-2}\,\mathrm{s}^{-1}$, which leads to an estimated neutrino luminosity $L_\nu \sim \int dE F_E\times 4\pi d_L^2\sim 2\times 10^{42}\,\mathrm{erg/s}$, similar to NGC~1068. We compare this with the proton luminosity obtained as a fraction $\mathcal{F}_p$ of the magnetic dissipation rate
\begin{equation}
    L_p\simeq 1.8\times 10^{43}\,\frac{\mathcal{F}_p \sigma_{\rm tur}^{3/2} R_{10}}{\eta}\,\mathrm{erg/s},
\end{equation}
which leads to the complementary constraint on the non-thermal proton energy budget
\begin{equation}
    \mathcal{F}_p\sigma_{\rm tur}^{1/2}\sim 10^{-2}.
\end{equation}
We see that the arguments leading to the conclusion that only a small fraction of the dissipated energy being transferred to non-thermal protons, $\mathcal{F}_p\ll 1$, are very simple and apparently unavoidable. The IceCube signal peaks at very large energies, favoring a strongly magnetized turbulent corona with a large magnetic dissipation rate, yet the normalization of the signal implies that a rather small fraction of the available energy is dissipated in non-thermal protons. Similarly, the fraction of energy going into X-rays in the 2-10~keV band is
\begin{equation}\label{eq:FX_7469}
    \mathcal{F}_X\simeq 0.07\frac{\eta}{\sigma_{\rm tur}^{3/2}R_{10}}\sim \frac{4\times 10^{-4}}{\sigma_{\rm tur}^{1/2}}.
\end{equation}
Clearly with such a small rate of energy dissipated in non-thermal particles, their impact on the turbulent cascade is negligible, with 
\begin{equation}
    \frac{U_{p,\rm nth}}{U_{B,\rm tur}}\simeq \frac{3\times 10^{-3}}{\sigma_{\rm tur}}.
\end{equation}
Even more than in the case of NGC~1068, this indicates that non-thermal proton feedback on the turbulent cascade is unimportant.

\section{Diffuse neutrino emission from Seyfert galaxies}\label{sec:diffuse}

\begin{figure*}[t]
    \centering
\includegraphics[width= \textwidth]{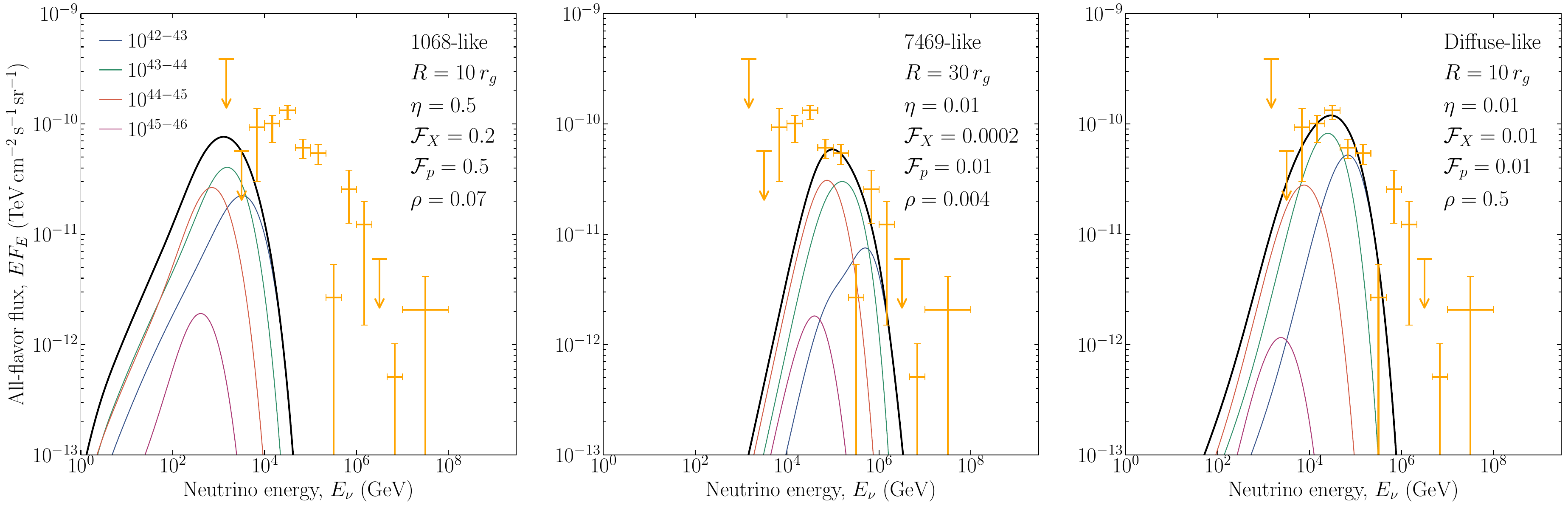}
    \caption{Predicted diffuse neutrino flux from a population of Seyfert galaxies with coronal parameters informed by our fit for NGC~1068 (left), NGC~7469 (center), and intermediate properties that fit by eye the spectral break of the diffuse flux (right). Different colors indicate the contribution of coronae of different $L_X^{2-10\,\rm keV}$, in $\rm erg\,s^{-1}$, as indicated in the legend. In all cases, we rescale the total flux by a factor $\rho<1$, to match the normalization of the diffuse flux observed by IceCube. Measured flux points are taken from \cite{IceCube:2025ewu}.}\label{fig:diffuse}
\end{figure*}

\begin{figure*}[t]
\includegraphics[width=\textwidth]{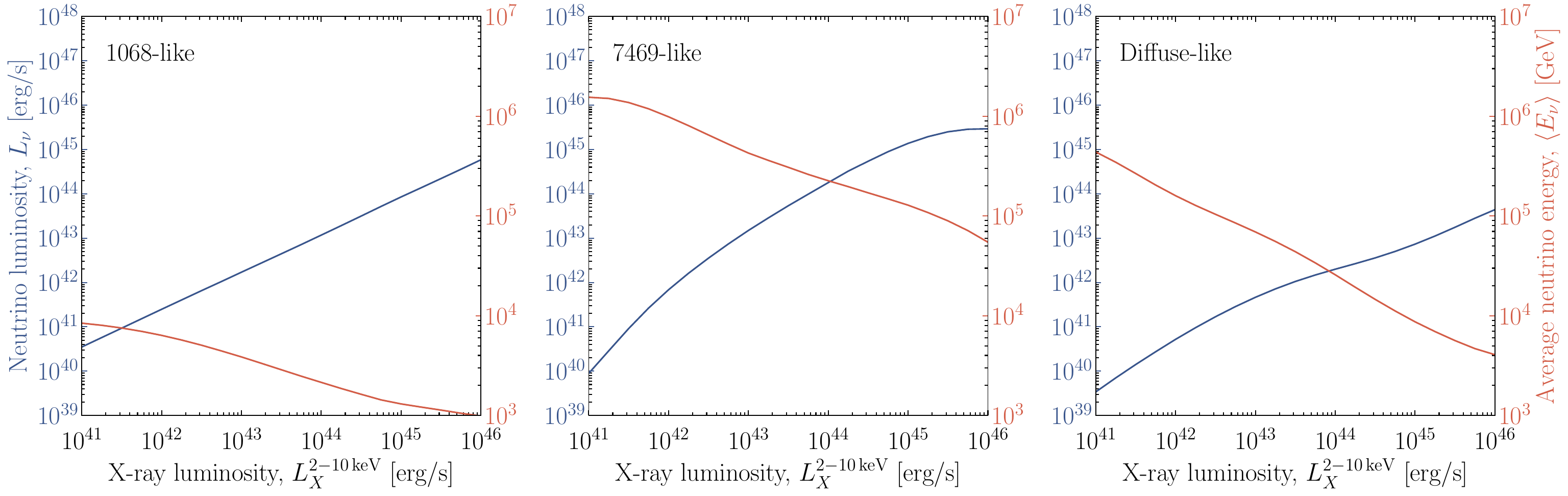}
\caption{Neutrino luminosity and characteristic peak neutrino energy (defined as the average energy over the energy distribution) for varying $L_X^{2-10\,\rm keV}$.}\label{fig:sequence}
    
\end{figure*}

Since individual AGN produce high-energy neutrinos, as inferred from the NGC~1068 case, and potentially from the other tentative excesses, one may naturally expect them to contribute to the diffuse neutrino flux. This expectation has even more grounds in view of the recent IceCube measurement of a spectral break in the diffuse neutrino flux around 20-30~TeV~\citep{IceCube:2025tgp}, roughly the region where most neutrino excesses from Seyfert galaxies seem to peak. Indeed, several works have followed this idea, using different modeling for the neutrino emission from individual AGN~\citep{Murase_2020,Ambrosone:2024zrf,Padovani:2024tgx,Fiorillo:2025ehn,Karavola:2026rpg,Murase:2026hrz}. In this section, we use the results of our analysis for NGC~1068 and NGC~7469 to estimate the AGN contribution to the diffuse neutrino flux within our turbulence-based scenario.

We stress that such an analysis cannot be conclusive or strongly predictive, due to the generally unknown properties of individual AGN coronae. For each galaxy, the parametric freedom of our model is encoded in the magnetization $\sigma_{\rm tur}$, coherence length $\eta$, radius $R$, and non-thermal proton energy budget $\mathcal{F}_p$; these four parameters could vary significantly across the entire galaxy population. Nevertheless, we can take two extreme scenarios, 1068-like and 7469-like, in which all coronae are assumed to have as parameters exactly the best-fit values we have inferred for NGC~1068 and NGC~7469, respectively. A more realistic situation, in which different AGN have different coronal properties, will likely interpolate among these two extremes.

We characterize the population of AGN in terms of their measured $L_X^{2-10\,\mathrm{keV}}$, the de-absorbed X-ray luminosity in the 2-10~$\mathrm{keV}$ band. Specifically, we parameterize the luminosity and redshift evolution of the AGN with the prescription of \cite{Ueda:2014tma}. From here, we extract the luminosity-dependent density evolution in terms of $d\Phi/d\log_{10}\tilde{L}_X^{{2-10 \,\mathrm{keV}}}$ (measured in $\mathrm{Mpc}^{-3}\,\mathrm{dex}^{-1}$), the number of sources per comoving volume per decade of the comoving intrinsic X-ray luminosity in the 2-10 $\mathrm{keV}$ range, $\tilde{L}_X^{2-10 \,\mathrm{keV}}$. We also assign to each AGN a target photon spectrum, with a dependence on their $L_X^{2-10\,\mathrm{keV}}$ that follows the sequence of~\cite{Marconi_2004_SED}. 

Beyond the X-ray luminosity, the other fundamental property of the corona is its magnetic field. So far, we have measured its strength in terms of the turbulent magnetization $\sigma_{\rm tur}$. On the other hand, if the corona is assumed to be magnetically powered, a complementary, and potentially more useful, measure is given by the ratio between the instrinsic X-ray luminosity $L_X$ and the magnetic dissipation rate $L_B$. For a magnetically powered corona, this ratio is expected to be  below 1. As discussed in Section~\ref{sec:multimessenger_emission}, the physical parameter that we use to describe the X-ray power output of each corona is $\mathcal{F}_X = L_X^{2-10\,\mathrm{keV}}/L_B$.


We also assume that the mass of the central black hole $M$, which in turn fixes the typical length scale $r_g$, scales directly with $L_X^{\rm 2-10\,keV}$, following the empirical relation shown in Fig.~11 of \cite{Mayers:2018hau}
\begin{equation}
    M = 2\times10^7\,\left(\frac{L_X^{2-10\,\mathrm{keV}}}{2\times10^{43}\,\mathrm{erg}\,\mathrm{s}^{-s}}\right)^{0.58}\,M_\odot\,.
    \label{eq:bh_mass}
\end{equation}
This allows us to express the gravitational radius in terms of the X-ray luminosity as
\begin{equation}\label{eq:radius_LX}
    r_g \simeq 3 \times 10^{12}\,\left(\frac{L_X^{\text{2-10}\,\mathrm{keV}}}{2\times 10^{43}\,\mathrm{erg}\,\mathrm{s}^{-1}}\right)^{0.58}\,\mathrm{cm}\,.
\end{equation}

We assume a constant $\mathcal{F}_X$, equal to the best-fit values for NGC~1068 and NGC~7469 respectively, so that for each galaxy we obtain the magnetization as
\begin{equation}\label{eq:sigma_diffuse}
    \sigma_{\rm tur}\simeq  \left(\frac{L_X^{2-10\,\mathrm{keV}}}{2\times 10^{43}\,\mathrm{erg/s}}\right)^{0.28}\,\left(\frac{\eta}{\mathcal{F}_XR/r_g}\right)^{2/3}.
\end{equation}

For both scenarios (1068-like and 7469-like), we follow the method described in Section 2 and use $\texttt{AM}^3$  to obtain $dN_\nu/dE_\nu dt = E_\nu^{-1}dL_\nu/dE_\nu$, the number of neutrinos emitted per unit time and energy in the rest frame of the source.
We integrate this quantity over the population of AGN using a standard-candle approximation, assuming that all the galaxies are described by the same intrinsic model parameters: the only change is $L_X^{2-10\,\mathrm{keV}}$ and correspondingly the length scale $r_g$, which scales with the SMBH mass. With these definitions, the diffuse neutrino flux at Earth is
\begin{equation}
    \frac{d\Phi_\nu}{dE_\nu} = \frac{c}{4\pi} \int_0^{+\infty}\frac{dz}{H(z)} \int d\tilde{L}_X^{{2-10}} \frac{d\Phi}{d\tilde{L}_X^{{2-10}}} \frac{dN_\nu}{dE_\nu\,dt}[E_\nu(1+z)]\,,
\end{equation}
where $H(z)=H_0 \sqrt{\Omega_\Lambda+\Omega_{\rm mat}(1+z)^3}$ is the Hubble parameter, in terms of its local value $H_0$ and of the energy fraction of dark energy $\Omega_\Lambda = 0.685$ and matter $\Omega_{\rm mat} = 0.315$. 

We find that assuming that all galaxies behave exactly like NGC~1068 or NGC~7469 leads to a significant overshoot of the diffuse neutrino flux. Therefore, we rescale the signal prediction by a fraction $\rho$, meant to model the fraction of all AGN which are neutrino emitters as efficient as our benchmark in each scenario.

Fig.~\ref{fig:diffuse} shows the resulting diffuse neutrino flux for both scenarios: we highlight in the figure the model parameters for each of them, including the fraction $\rho$ by which we have to rescale to match the observed signal. We do not include in these results galaxies with $L_X^{2-10\,\mathrm{keV}}<10^{42}\,\rm erg\,s^{-1}$, to which the parameterized SED from~\cite{Marconi_2004_SED} may not directly apply due to the limited statistics for such faint galaxies in that work. Assuming a constant $\mathcal{F}_X$ as we do, their contribution to the diffuse flux is anyway negligible, since their low luminosity implies a comparably low magnetic dissipation rate. 

For the NGC~1068 benchmark, the diffuse neutrino flux is rather similar to the one obtained in~\cite{Fiorillo:2025ehn}. The main qualitative difference is that it peaks at even lower energies: this is driven by our fit to the NGC~1068 neutrino flux peaking around 1~TeV, whereas~\cite{Fiorillo:2025ehn} was based upon a fit ``by eye'' to the signal peaking at around 10~TeV. On the other hand, we agree with~\cite{Fiorillo:2025ehn} on other relevant accounts: the overall signal is dominated by galaxies with X-ray luminosity similar to NGC~1068, due to the rarity of higher-luminosity objects, and only a fraction $\rho\sim 0.07$ must be efficient in neutrino emission to avoid overshooting the diffuse flux. It is also worth noting that the requirement of about $10\%$ of galaxies being as efficient as NGC~1068 is somewhat more general than the turbulent model. An independent analysis based on the reconnection model~\citep{Karavola:2026rpg} recovers roughly the same fraction.

For the scenario based on NGC~7469, the properties are, as expected, very different. The diffuse neutrino flux peaks in this case at around $100\,\mathrm{TeV}$, again dominated by galaxies with X-ray luminosity in the range of NGC 1068 and NGC 7469. In this scenario, the fraction of galaxies needed to power the diffuse flux at 100~TeV is very low, $\rho\sim 0.004$. This value should be considered as an upper limit for this idealized scenario. Indeed, from Fig.~\ref{fig:diffuse} we see that the galaxies contributing to the diffuse at $\sim100\,\rm TeV$ are those with $L_X \sim 10^{43}\,\rm erg\,s^{-1}$, and from Fig.~\ref{fig:sequence} we see that those would emit a neutrino luminosity of $L_\nu\sim10^{43}\,\rm erg\,s^{-1}$, which is similar to the one with which NGC 1068 was discovered. Furthermore, this neutrino signal would be produced in the energy range of $\sim 100\,\rm TeV$, in which the atmospheric background is even lower, facilitating a potential detection. For these reasons, this scenario seems to be in tension with the discovery of only one source with these properties, NGC 7469. However, to make such a statement more robust, one would need to perform a full statistical comparison of the expected number of neutrino coincidences with individual sources, which we plan to perform in a future work. Moreover, in this scenario the turbulent magnetization obtained by Eq.\ref{eq:sigma_diffuse} reaches values of $\sigma_{\rm tur}\ge5$ for galaxies with $L_X^{2-10\,\mathrm{keV}}>10^{45}\,\rm erg\,s^{-1}$, where acceleration driven by magnetic reconnection could play a significant role; on the other hand, these galaxies make anyway a negligible contribution to the diffuse flux, due to their low neutrino peak energies caused by the strong photohadronic cooling.

Both the 1068-like and the 7469-like scenarios cannot directly explain the spectral break observed by IceCube. It is rather simple to understand this conclusion, descending from the mismatch between the spectral break at 20-30~TeV, the NGC~1068 signal peaking at 1~TeV, and the NGC~7469 signal, whose best-fit explanation requires a neutrino spectrum peaking at 100~TeV. Therefore, if the diffuse neutrino at the spectral break is indeed powered by AGN coronae, their maximum proton energies should be somewhat intermediate between these two cases. As an example, we show the diffuse neutrino flux predicted for a third scenario, where a high fraction $\rho \sim 0.5$ of the AGN emit neutrinos from a corona of radius $R=10\,r_g$, such as the one of NGC 1068, but with a lower value of $F_X$ and $\eta$ chosen so that the neutrino peak matches the spectral break in the diffuse flux. This diffuse-like scenario is shown in the third panel of Fig.~\ref{fig:diffuse}, reproducing closely the flux measured by IceCube close to its peak. 

For all of these scenarios, even though the coronal parameters are assumed identical for all AGN coronae, the neutrino output depends in a non-linear way on the X-ray luminosity. We already stressed this point in~\cite{Fiorillo:2025ehn}; with our newly favored scenarios, we show the dependence of the neutrino luminosity $L_\nu$ and average neutrino energy $\langle E_\nu\rangle$ on the intrinsic X-ray luminosity in the $2-10\,\rm keV$ band in Fig.~\ref{fig:sequence}. 

We find again, as in~\cite{Fiorillo:2025ehn}, that the neutrino luminosity scales sub-linearly with the X-ray luminosity, due to the increased cooling in the brightest galaxies. The specific dependence of $L_\nu$ on $L_{X}^{2-10\,\mathrm{keV}}$ changes significantly among the three scenarios, and in the 7469-like scenario in particular does not even approximately follow a power law, with $L_\nu$ saturating at high luminosities. This non-linear trend is expected, since the efficiency of neutrino production depends on a relative balance between acceleration and photohadronic cooling. For the average neutrino energy, increasing $L_X^{2-10,\rm keV}$ always leads to a lower typical neutrino energy, due to the more intense photohadronic cooling that limits the maximum energies that protons can achieve. The dependence of $\langle E_\nu\rangle$ on the luminosity is less pronounced than that of $L_\nu$, with $\langle E_\nu\rangle$ changing by slightly more than one order of magnitude over the entire range of $L_X^{2-10\,\rm keV}$. Overall, these results serve as a clear warning against using fixed spectral templates for all Seyfert galaxies rescaled using simple linear prescriptions: in our model, both the luminosity $L_\nu$ scaling, and the average neutrino energy $\langle E_\nu \rangle$, can change in highly non-linear ways with the galaxy luminosity, depending on the model parameters. For this reason, we believe our GitHub release~\citep{GitRepo} provides a much more reliable and comprehensive tool to explore the model predictions for neutrino emission from a population of Seyfert galaxies.

Clearly none of the scenarios we study, based on identical intrinsic parameters for all AGN coronae, is realistic. If both NGC~1068 and NGC~7469 are interpreted as real neutrino emitters, then the most likely conclusion is that there is significant variance of coronal properties among the different sources, and therefore the true diffuse neutrino flux would presumably interpolate between these ranges. It is intriguing to speculate that such an interpolation, with a small fraction of AGN being very efficient accelerators as NGC~7469, and a larger fraction peaking at lower neutrino energies, might explain all of the diffuse neutrino flux between 1~TeV and 1~PeV. Still, one should keep in mind that achieving neutrinos at hundreds of TeV within the turbulent coronal model requires somewhat tuned parameters, as we have seen in the case of NGC~7469, and as we discuss in more detail in Secs.~\ref{sec:back_of_the_envelope} and~\ref{sec:discussion}. An alternative possibility is that the highest-energy neutrinos, above 100~TeV, may be powered by low-luminosity AGN, with $L_X\lesssim 10^{41}\,\mathrm{erg/s}$: in our Figs.~\ref{fig:diffuse}, their contribution is suppressed due to our assumption of a constant $\mathcal{F}_X$, which implies that dimmer AGN are less magnetized. If this assumption is relieved, then low-luminosity AGN, which are dimmer and therefore have longer cooling timescales for protons, might contribute at higher energies. On the other hand, the SED extracted from \cite{Marconi_2004_SED} was obtained only for brighter AGN, and therefore extrapolating it to lower luminosities may not be justified, which is why we do not attempt such an explanation in more detail in this work.


\section{Discussion}\label{sec:discussion}

Non-thermal protons accelerated by strong turbulence ($\delta B \sim B$) in a magnetized coronal environment are a viable explanation for the neutrino excesses from Seyfert galaxies reported by IceCube. In our previous works~\citep{Fiorillo:2024akm,Fiorillo:2025ehn}, we based this conclusion upon a PIC-inspired model of particle acceleration in strongly magnetized turbulence, showing that large magnetizations $\sigma_{\rm tur}\sim 1$ are required to explain the neutrino luminosity reported by IceCube from NGC~1068. Here, we have extended this result by accounting for the full multimessenger emission, including the electromagnetic signal. Through the resulting emissivity, we have obtained for the first time explicit constraints from the combined IceCube and Fermi-LAT measurements on the magnetic properties of the turbulence, parameterized by the turbulent magnetization $\sigma_{\rm tur}$, the dimensionless coherence length $\eta$, the coronal size $R$, and the fraction of dissipated magnetic energy in non-thermal protons $\mathcal{F}_p$. Our results are also supported by simple analytical estimates, and point to an intriguing tension between NGC~1068 and NGC~7469, the currently most globally significant cases.

In the main text, we have focused on the two galaxies from which IceCube has reported the highest post-trial significance~\citep{Abbasi:2025tas}. For NGC~1068, we find a solution that favors a strongly magnetized ($\sigma_{\rm tur}\sim 1$), compact corona ($R\sim 10\,r_g$) with relatively large coherence length ($\eta = \ell/R\sim 0.5$) and with a large fraction of energy dissipated in non-thermal protons ($\mathcal{F}_p\sim 0.5$). This ultimately comes from the bright neutrino signal peaking at a relatively low energy $E_\nu\sim 1\,\mathrm{TeV}$, making the galaxy a powerful accelerator reaching only to relatively low neutrino energies. 

A somewhat opposite result is reached for NGC~7469, which is associated with two neutrino events at hundreds of TeV. This favors a very efficient accelerator with only a small fraction of power going in non-thermal protons, requiring a small coherence length ($\eta\sim 0.01$), and a slightly larger coronal radius ($R\sim 30\,r_g$) to dilute the photon field from the accretion disk and limit the impact of cooling. The fraction of energy dissipated in non-thermal protons is small ($\mathcal{F}_p\sim 0.01$), but a more concerning conclusion is that the power in the $2-10\,\rm keV$ X-ray band is even smaller ($\mathcal{F}_X\sim 2\times 10^{-4}$). Such an estimate suggests that the required magnetic dissipation power $L_B$ may even exceed the accretion power of the AGN. This conclusion is largely caused by the large magnetic fields required to accelerate protons up to the very high energies consistent with the IceCube signal. Our back-of-the-envelope analysis shows (see Eq.~\ref{eq:FX_7469}) that this tension can be relieved only at the cost of assuming a very small magnetization $\sigma_{\rm tur}\lesssim 10^{-2}$, but then the condition to achieve neutrinos at 100~TeV (Eq.~\ref{eq:first_constraint_7469}) would imply extremely small coherence length for the turbulence $\eta\lesssim 10^{-4}$. So the rather large neutrino energy signaled from NGC~7469 raises some tensions with the model, an interesting aspect that might be confirmed or disproved by future neutrino measurements. 

For the other two galaxies considered in this work, NGC 4151 and CGCG 420-015, the signal is not significant enough, and the constraints in the parameter space, reported in Appendix~\ref{appendix:altre}, are not conclusive. One intriguing conclusion regarding NGC~4151 is that the predicted gamma rays associated with the neutrino emission seem to follow closely the high-energy Fermi-LAT observations, raising the possibility that we are indeed observing coronal gamma rays, as shown in Fig.~\ref{fig:A1}. This conclusion had also been reached by \cite{Murase:2023ccp}, using a turbulence model with a somewhat weaker magnetization, with a plasma beta parameter $\beta=1$: for comparison, our best-fit solution reported in Table~\ref{tab:best-fit} has a beta parameter more than an order of magnitude lower. 

This is part of a general trend maintained also for other galaxies, where in comparison to other works based on turbulence scenarios we find larger magnetic fields and smaller plasma beta parameters. A similar difference is found, for NGC~1068, with the inference from recent work by~\cite{Eichmann:2026kvj} favoring values of $\beta\gtrsim 1$. We trace the difference, in comparison with other works from the literature, to our adoption of the constraint $L_p\lesssim L_B$, which requires sufficiently strong magnetic fields to achieve neutrino luminosities comparable with the IceCube excess. Removing this constraint would require some additional non-magnetic form of dissipation.

Our analysis of the IceCube data is based upon the reconstruction of a mock data sample through the public response functions in the $\texttt{PLE}\nu\texttt{M}$ code~\citep{Schumacher:2025qca}, since the observations associated with the recent IceCube works on Seyfert galaxies~\citep{Abbasi:2025tas} are not yet public. In this sense, an analysis from the Collaboration itself, or using the actual IceCube data, might in principle reveal more stringent constraints. For this reason, we have made the full parametric predictions of our turbulent model public in a GitHub release, spanning both electromagnetic and neutrino emission~\citep{GitRepo}.
An additional messenger, which might shed in the future more light on the nature of coronal emission, is the radio band. As first proposed by \cite{Inoue_2020}, the radio excess observed from NGC~1068 in the millimiter band might plausibly be of coronal origin. At present, we refrain from comparing our turbulent scenario predictions with these measurements, since the low-energy cascaded emission from the corona presumably depends upon the primary non-thermal electrons, whose distribution may be strongly anisotropic in pitch angle \citep{Comisso:2019frj,CS21,Com2022}, as well as the re-acceleration of electrons involved in the electromagnetic cascade, which we are currently not accounting for. This is therefore a natural avenue for further exploration from the modeling perspective. Another topic which we plan to examine in upcoming work is a systematic population study to assess the expected number of IceCube associations with known Seyfert galaxies. Such a study would necessarily be based on assumptions on the statistical spread of parameters among different AGN, yet it would be a powerful test of the model itself.

For all the best-fit solutions identified in this work, the non-thermal proton energy density is a small fraction of the magnetic energy density, $U_{p,\rm nth}\lesssim 0.1 U_{B,\rm tur}$, as shown in Table~\ref{tab:best-fit}. This suggests that the overall impact of proton acceleration on the turbulent cascade might be small, although solutions in the opposite regime also exist~\citep{Lemoine:2025roa}. In our model, the small ratio between proton and magnetic energy density is largely dictated by the constraint that the energy dissipated per unit time in protons does not exceed the magnetic dissipation rate, i.e. that $\mathcal{F}_p<1$, as explained in Sec.~\ref{sec:energetics}.

We have also obtained, for both scenarios inspired by NGC~1068 and NGC~7469, the diffuse neutrino flux, establishing that only a small fraction of galaxies of $7\%$ and $0.4\%$, respectively, can be as efficient as these two reference cases in order not to exceed the diffuse neutrino flux observed by IceCube. Given the widely different neutrino signals reported from these two galaxies, the diffuse fluxes predicted in the two scenarios are also very different. The diffuse neutrino flux in either scenarios is incompatible with the spectral break observed by IceCube, which lies at around 20-30~TeV, too high in comparison with the peak of the neutrino spectrum from NGC~1068, and too low in comparison with the 100-TeV neutrinos from NGC~7469. This suggests that, if Seyfert galaxies are powering the diffuse neutrino flux around the spectral break, the majority of them have properties somewhat intermediate, in terms of magnetization and coherence scale of the turbulence, between those inferred for these two sources.


Overall, we have shown that neutrino observations, in combination with gamma-ray upper limits, impose stringent bounds on the intrinsic properties of strongly magnetized turbulence in AGN coronae --- bounds that are largely inaccessible to purely electromagnetic probes. As more data are collected from the Seyfert galaxies exhibiting tentative neutrino excesses, these bounds will progressively sharpen, opening the door to a quantitative and comparative understanding of coronal turbulence across different AGN systems.

\section*{Note added in preparation}

While this work was being finalized, ~\cite{Carpio:2026xkf} appeared which also uses the combined neutrino and gamma-ray observations to deduce constraints on a turbulent corona model. Our work differs primarily in the use of the non-resonant turbulent model developed in~\cite{Fiorillo:2024akm}, with an acceleration rate directly extracted from PIC simulations~\citep{Comisso:2019frj}. Furthermore,~\cite{Carpio:2026xkf} focuses primarily on the hadronic properties of the corona, whereas our work primarily considers the magnetic properties, informed also by the theoretical constraints on the proton energization rate being consistent with the magnetic dissipation rate. Our works also differ in the statistical treatment of the IceCube data.

\section*{Acknowledgements} 
We thank Mahmoud Al-Awashra for useful comments on this manuscript.
D.F.G.F. was supported by the Alexander von Humboldt Foundation (Germany) for most of the completion of the project.
L.C is supported by the NSF grant PHY-2308944 and the NASA ATP grant 80NSSC24K1230.
EP acknowledges support from INAF through ``Assegni di ricerca per progetti di ricerca relativi a CTA e precursori'' and from the Agence Nationale de la Recherche (grant ANR-21-CE31-0028).
L.S. acknowledges support from DoE Early Career Award DE-SC0023015, NASA ATP 80NSSC24K1238, NASA ATP 80NSSC24K1826, and NSF AST-2307202. This work was supported by a grant from the Simons Foundation (MP-SCMPS-00001470) to L.S. and facilitated by Multimessenger Plasma Physics Center (MPPC), grant NSF PHY-2206609 to L.S.

\bibliographystyle{aasjournal}
\bibliography{References}

\appendix

\begin{table*}[t]
\centering
\begin{adjustbox}{max width=\textwidth}
\begin{tabular}{|l|c|c|c|c|c|c|}
\hline
Source name & $\Phi_{\nu_\mu + \bar{\nu}_\mu}^{1\, \rm TeV}$ $(\rm TeV^{-1}\,cm^{-2}\,s^{-1})$ & $\gamma$ & $E_{\nu, \,\rm end}$ $(\rm GeV)$ & $\rm Dec \,(deg)$& $n_{\rm sig}^{\rm IC}$ & $\tilde{n}_{\rm sig}$  \\
\hline
NGC 1068 & 4.7$\times10^{-11}$ & 3.4 & $2\times 10^4$ & -0.01 & 102.2 & 119.7 \\
\hline
NGC 7469 & 2$\times10^{-13}$ & 1.9 & $1.8\times 10^6$ & 8.87 & 5.5 & 3.6 \\
\hline
NGC 4151 & 6$\times10^{-12}$ & 2.7 & $3.5\times 10^4$ & 39.41 & 27.6 & 27.4 \\
\hline
CGCG 420-015 & 8.7$\times10^{-12}$ & 2.7 & $2.5\times 10^4$ & 4.06 & 35.3 & 34.1 \\
\hline
\end{tabular}
\end{adjustbox}
\caption{Parameters used for our analysis with \plenum. Power-law parameters are the best-fit values reported in \cite{Abbasi:2025tas, Bellenghi:2024dcp}, together with the corresponding number of signal events $n_{\rm sig}^{\rm IC}$ reconstructed in the IceCube analysis. $\tilde{n}_{\rm sig}$ is the number of signal events reconstructed with our method for an exposure time of 13.1 years.  }
\label{tab:powerlaw}
\end{table*}

\section{Reproduction of IceCube neutrino signal for Seyfert galaxies}
\label{appendix:power-law}

Since the IceCube data relative to the recently reported excess from the four Seyfert galaxies analyzed in this study are not public, we do not perform an analysis on the actual data, but rather on mock event counts retrieved from~\cite{Abbasi:2025tas} through publicly available resources. To test the potential impact of this procedure, in this appendix we proceed to analyze the mock data sample with the same power-law spectrum used in the IceCube analyses.

The flux we use to produce the event distribution with \plenum\, is a power law parameterized, in units of $\rm TeV^{-1}\,cm^{-2}\,s^{-1}$, as
\begin{equation}
    \label{eq:PL_flux}
    \phi_{\nu_\mu+\bar{\nu}_\mu} = \Phi_{\nu_\mu+\bar{\nu}_\mu}^{1 \rm TeV} \left(\frac{E_\nu}{\rm TeV}\right)^{-\gamma}\,.
\end{equation}
As explained in Sec.~\ref{sec:statistical_analysis}, the mock dataset that we use as \textit{true} data is the event distribution computed by \plenum\, assuming a superposition of the background neutrino flux and
the signal flux. For the signal, we simulate the flux of Eq.~\ref{eq:PL_flux} up to the highest value of the energy range reported by IceCube, $E_{\nu, \,\rm end}$, and setting the model parameters to the best-fit values. These parameters are listed in Table~\ref{tab:powerlaw}. Following the procedure of Sec.~\ref{sec:statistical_analysis}, we compute the $\chi^2$ assuming the model power-law flux and minimize it to obtain the allowed regions in the two-dimensional parameter space. Our analysis successfully reconstructs for each source the assumed parameters as the best-fit values as expected. We also report in Table~\ref{tab:powerlaw} the number of signal muon track events predicted at IceCube $\tilde{n}_{\rm sig}$ from our pipeline, compared with the number of signal muon track events reported by IceCube $n_{\rm sig}^{\rm IC}$. We find reasonably close results, although our pipeline systematically overpredicts the number of signal events.

\begin{figure*}[t!]
    \includegraphics[width=\textwidth]{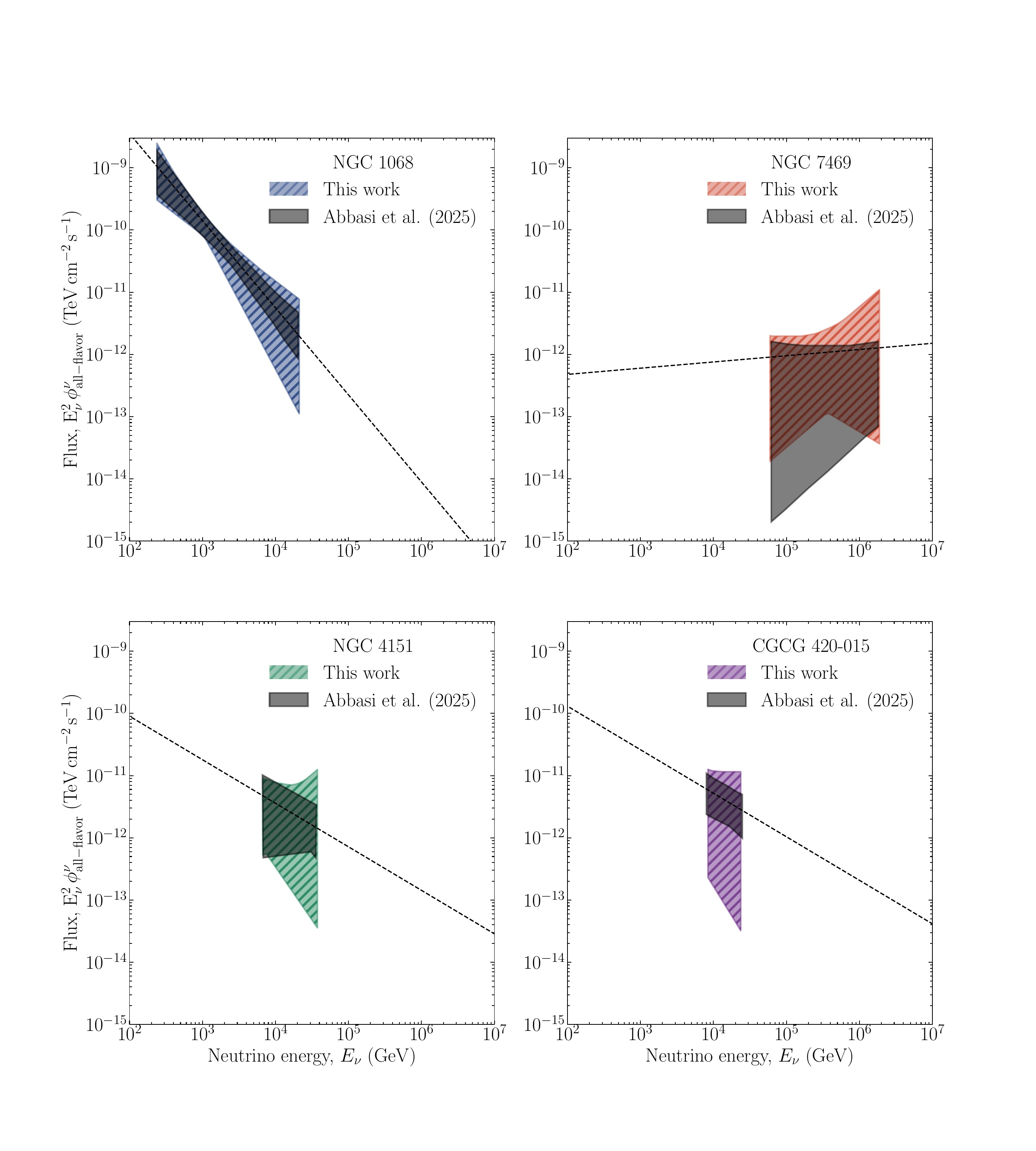}
    \caption{Comparison between the all-flavor power-law butterflies reported by IceCube \citep{Abbasi:2025tas} and those reconstructed with our pipeline. We plot the allowed power-law fluxes of Eq.~\ref{eq:PL_flux} allowed in the 68\% C.L. in the $\Phi_{\nu_\mu+\bar{\nu}_\mu}^{1 \rm TeV} -\gamma$ plane. For visualization clarity we plot them on the same energy range reported in \cite{Abbasi:2025tas}, although this pipeline leads to significantly different results in the energy range of the signal.
    }\label{fig:butterfly}
    
\end{figure*}

In Fig.~\ref{fig:butterfly}, we show for each source the all-flavor flux $\phi_{\rm all-flavor} = 3\phi_{\nu_\mu+\bar{\nu}_\mu}$ and compare the power-law uncertainty intervals for the flux reconstructed with our pipeline with those reported by IceCube~\citep{Abbasi:2025tas}. Overall, our method is able to reproduce qualitatively the results of the IceCube Collaboration. However, the lack of publicly available data and the limited resolution of the available response functions introduce intrinsic limitations to the accuracy of our analysis. This limitation applies especially to the energy range where the neutrino flux is shown. We choose to show the flux in the same energy range as~\cite{Abbasi:2025tas}, where this range is obtained as the one where most of the total TS of the analysis is collected. On the other hand, according to this definition, we are unable to reproduce the energy range using the dedicated notebook in \plenum\footnote{The notebook is available on   \href{https://github.com/PLEnuM-group/Plenum/blob/main/notebooks/misc/energy_range.ipynb}{github}}, which systematically yields broader energy ranges.
Finally, we note that the fact that our reconstructed butterflies are systematically less constrained than the one reported by IceCube is a good hint that a fit of our physical model to the data, with the analysis performed in \cite{Abbasi:2025tas}, would yield more stringent constraints that the ones of this work.

\section{Analysis of NGC 4151 and CGCG 420-015}
\label{appendix:altre}

In this section, we present the results of our statistical analysis as applied to mock data based on the IceCube excess from NGC 4151 and CGCG 420-015. For these two galaxies, IceCube has reported excesses with a local significance of $3.1\,\sigma$ (NGC 4151) and $2.7\,\sigma$ (CGCG 420-015), but no globally significant excesses. Nevertheless, for completeness, we investigate the model predictions for these cases.
\begin{figure*}[t]
    \centering
    \includegraphics[width= \textwidth]{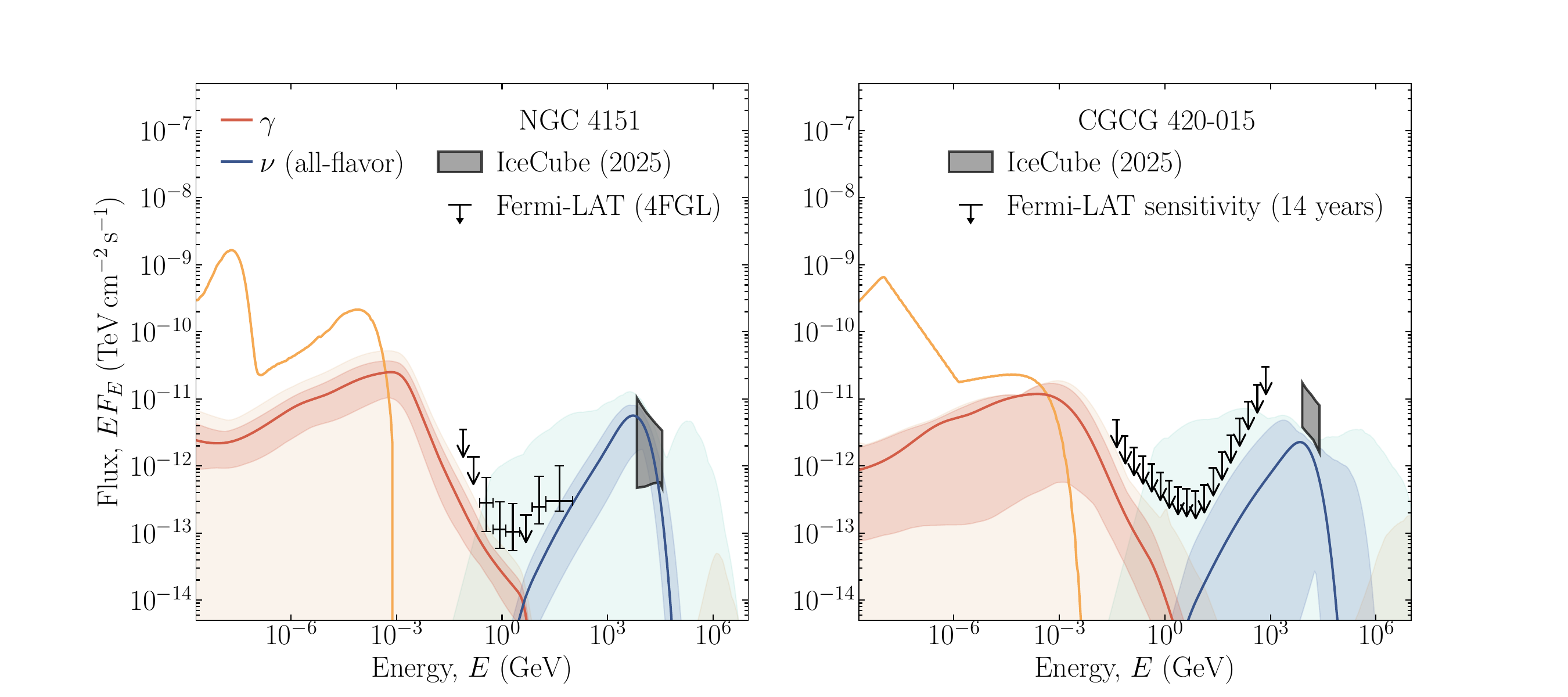}
    \caption{Neutrino and electromagnetic emission from the coronal region, for NGC~4151 (left) and CGCG~420-015 (right). We show the $68\%$ and $95\%$ confidence level (C.L.) regions with different shadings, together with the predicted best-fit curves. Since the confidence regions are shown mostly as a visual guide, we use the $\chi^2$ threshold for the TS corresponding to one degree of freedom, so as to match the exclusion contours for the individual parameters shown in Figs.~\ref{fig:4151_param} and~\ref{fig:CGCG_param}. The assumed SED is shown in orange for both galaxies. We do not include extragalactic gamma-ray absorption, which would only affect very-high-energy gamma rays above hundreds of GeV.}
    \label{fig:A1}
\end{figure*}

\begin{figure*}[t!]
    \includegraphics[width=\textwidth]{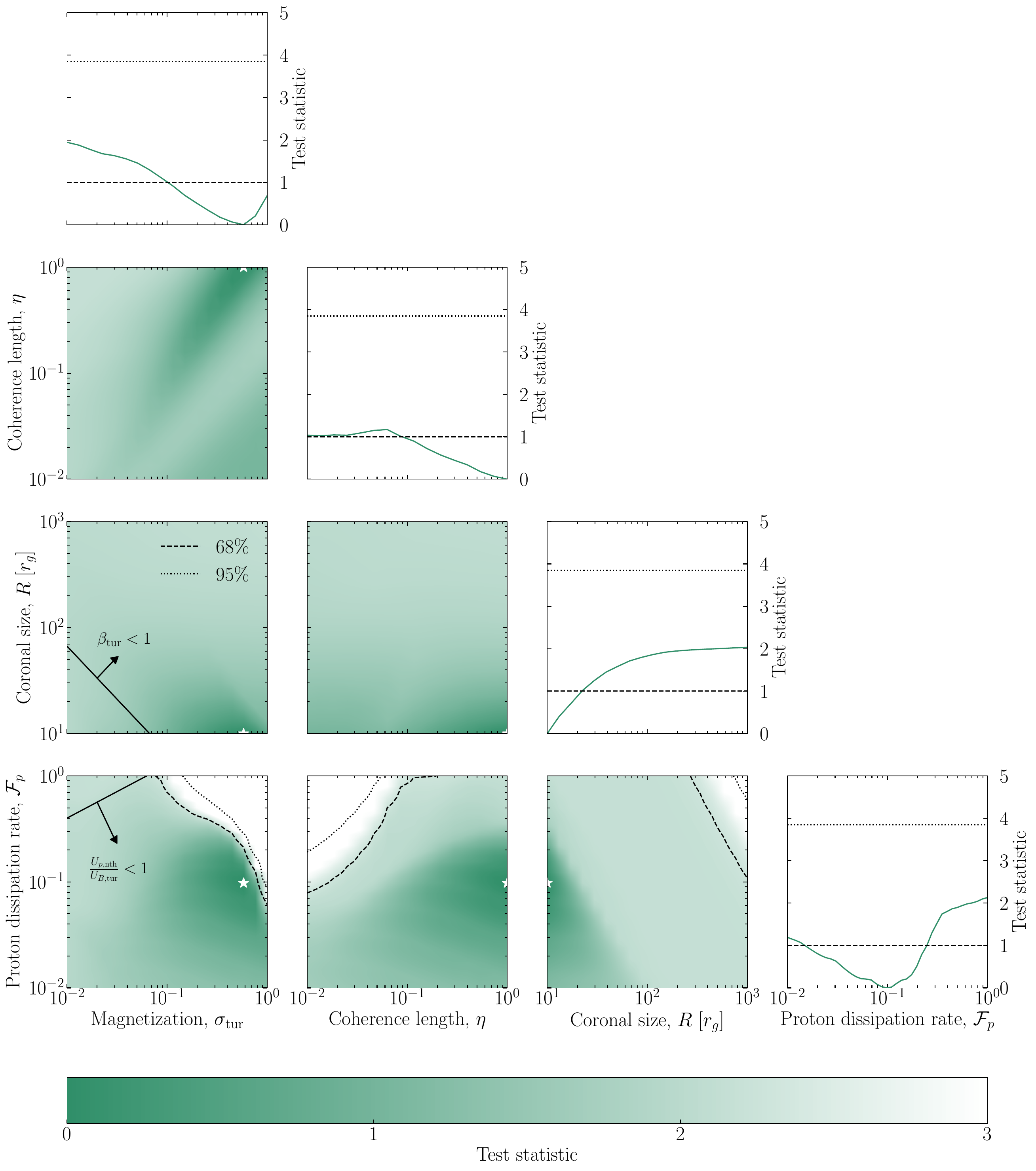}
    \caption{Constraints on coronal properties of NGC~4151 from the combined neutrino and gamma-ray signal. We highlight the regions of parameter space where $\beta_{\rm tur}<1$, signaling a strongly magnetized plasma, and $U_{p,\rm nth}/U_{B,\rm tur}<1$, signaling a small feedback of non-thermal protons on turbulence.}\label{fig:4151_param}
    
\end{figure*}

\begin{figure*}[t!]
    \includegraphics[width=\textwidth]{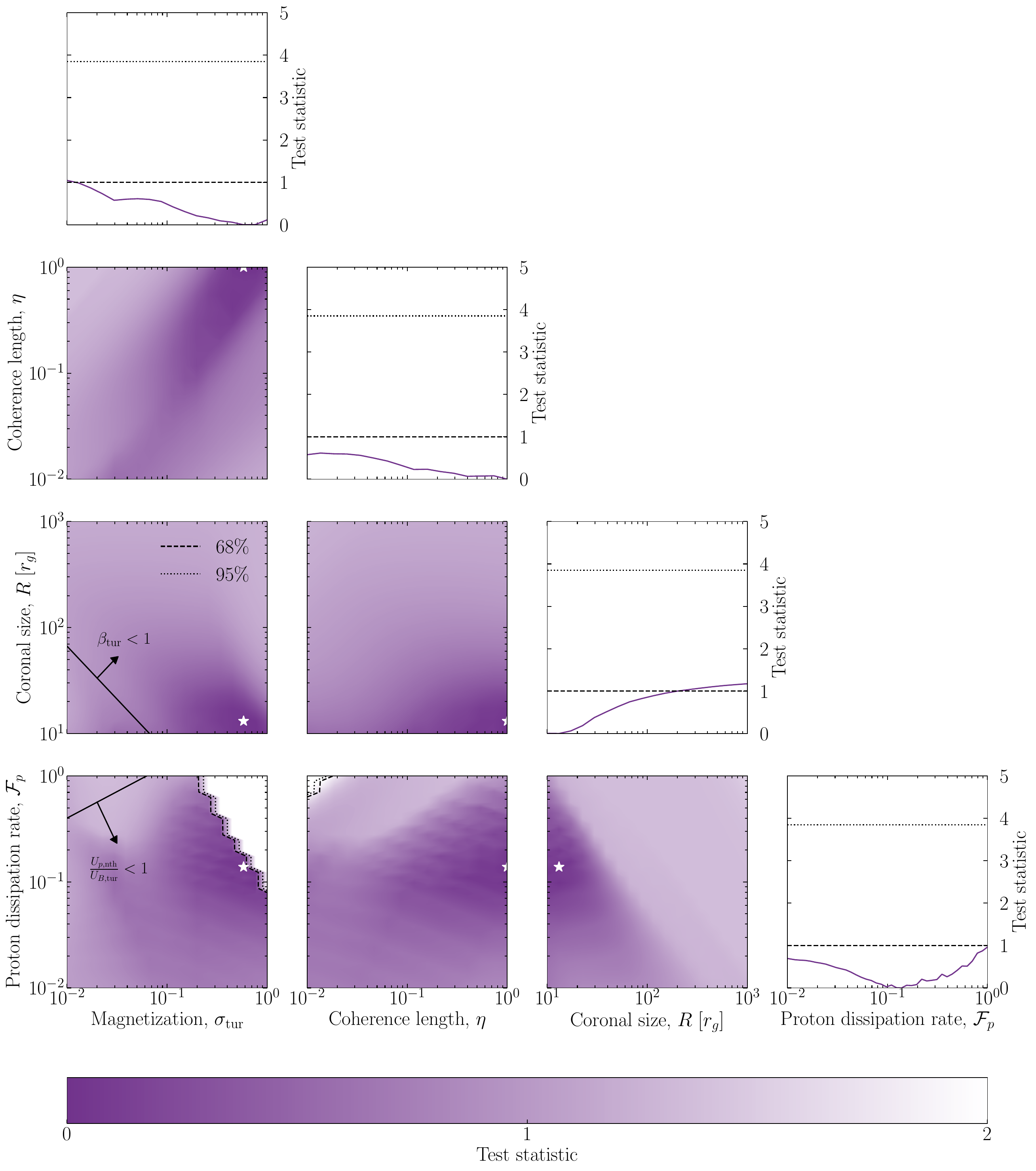}
    \caption{Constraints on coronal properties of CGCG~420-015 from the combined neutrino and gamma-ray signal. We highlight the regions of parameter space where $\beta_{\rm tur}<1$, signaling a strongly magnetized plasma, and $U_{p,\rm nth}/U_{B,\rm tur}<1$, signaling a small feedback of non-thermal protons on turbulence.}\label{fig:CGCG_param}
    
\end{figure*}
The tentative neutrino excess from NGC 4151 is reported at energies $\sim 10\,\rm TeV$, suggesting an acceleration efficiency somehow intermediate between the one of NGC 1068 and NGC 7469, that we discuss in the main text.
The gamma-ray emission from NGC 4151 is partially affected by source confusion. In particular, the 4FGL J1210.3+3928 source detected by the Fermi-LAT \citep{Fermi-LAT4FGL-DR3} is spatially consistent with both the blazar 1E 1217.9+3945 and NGC 4151. These two objects are separated by less than 5 arcminutes, a distance significantly smaller than the Fermi-LAT point-spread function at energies below 10 GeV, where the bulk of the emission is detected. 
In \citet{Murase_2024}, the entire gamma-ray signal was attributed to 1E 1217.9+3945, and upper limits were consequently derived for NGC 4151. Conversely, \citet{2025JCAP...07..013P} argued that the blazar is unlikely to dominate the emission below 10 GeV, which could instead be associated with NGC 4151, while a non-negligible contribution from the blazar may contaminate the observed spectrum above 10 GeV.
Given this ongoing debate regarding the correct flux attribution in the Fermi band, we adopt an agnostic approach and consider the flux reported for 4FGL J1210.3+3928 as an upper bound to the gamma-ray emission potentially associated with NGC 4151. Accordingly, we treat these data as the maximum flux level attainable by the electromagnetic cascade in our modeling.
With these assumptions, we performed the combined fit described in Section \ref{sec:analysis}. The favored solutions are shown in Fig.~\ref{fig:A1}, and indeed point to a neutrino flux peaking at $\sim10\,\rm TeV$. Interestingly, the allowed solutions at 68\% C.L. seem to indicate that the gamma rays of energies $10\,\rm MeV - 1\,\rm GeV$ produced in the corona, which in this case are the electromagnetic cascade of the hadronic gamma rays produced alongside neutrinos, can contribute to the signal detected by Fermi-LAT in this direction. Given the small local significance of the source as a neutrino emitter, the signal is not distinguishable from background at 95\% C.L.. For this reason, the constraints on the model parameters obtained for this source, shown in Fig.~\ref{fig:4151_param}, are not sufficient to exclude any value at $\ge 2\sigma$. We also note that some of the allowed solutions at 95\% C.L. peak at energies of $\sim 100 \,\rm TeV$. This is due to the fact that the energy resolution of IceCube, which is approximated in \plenum, becomes poorer at higher energies, so that a signal at $\sim 100 \,\rm TeV$ with this significance is hardly distinguishable from a signal at $\sim 10 \,\rm TeV$. The best-fit parameters, reported in Table \ref{tab:best-fit}, show that the favored dimensions of the corona are similar to the one of NGC 1068, i.e., $R\sim 10\,r_g$, but that the turbulent magnetization is lower and the coherence length is higher. In spite of this, the peak neutrino energies are higher than for NGC~1068, mostly because NGC 4151 is less luminous and has a higher black hole mass, which implies a larger length scale $r_g$. These two factors contribute to a weaker cooling on the accelerated protons, for a given $R/r_g$, compared to NGC 1068, so that even a longer $t_{\rm acc}$ is sufficient to accelerate protons at the $\sim 200\,\rm TeV$ required for the production of $\sim 10\,\rm TeV$ neutrinos.


Similar conclusions are drawn for CGCG 420-015. The signal for this galaxy is reported at energies of $\sim10\,\rm TeV$,
but for this source no Fermi analysis was available in the literature. For this reason, in place of a fiducial upper limit, we used the 14 years sensitivity at 5 sigma at a galactic latitude of $30^\circ$: a gamma-ray analysis of this source would therefore provide more reliable constraints. The best-fit solutions for this source had a significant cascade that in most cases was higher than the target SED. The requirement that the electromagnetic emission does not exceed the target is an additional constraint that we add to the fit, to ensure the self-consistency of our approach. The SEDs shown in Fig.~\ref{fig:A1} fall right outside the butterfly reported by IceCube, suggesting that in order to produce neutrinos at the level of luminosity reported, a cascade signal of comparable luminosity has to be expected. Also for this source we show the constraints on the parameter space in Fig.~\ref{fig:CGCG_param}, and similar to NGC 4151, the signal is not sufficient to exclude any parameter at a significance of $\ge 2 \,\sigma$. On the $\sigma_{\rm tur}-\mathcal{F}_p$ plane the effect of our additional constraint that the electromagnetic cascade must be smaller than the target SED is visible for high values of $\sigma_{\rm tur}$ and $\mathcal{F}_p$, for which the cascade would not respect this condition. Also for this source, the best-fit parameters suggest that a lower acceleration efficiency is needed to produce neutrinos at $\sim 10\,\rm TeV$. While the source is the most luminous among the four considered in this work, the mass of its SMBH is almost two orders of magnitude larger than the one of NGC 1068, so that a radius $R = 13 \,r_g$ is sufficient to dilute the photon field and allow acceleration to $\sim 100 \, \rm TeV$ even with values of $\sigma_{\rm tur} = 0.6$ and $\eta =1$.
\end{document}